\documentclass{article}
\usepackage{a4wide,epsfig}
\usepackage{amsmath,amssymb}
\usepackage{natbib}

\begin{document}

\title{A maximum principle for the mutation-selection\\ 
equilibrium of nucleotide sequences}
\author{Tini Garske and Uwe Grimm\\[2ex]
Applied Mathematics Department, 
Faculty of Mathematics and Computing,\\
The Open University, Walton Hall,
Milton Keynes MK7 6AA, UK}
\date{\today}
\maketitle

\begin{abstract}
We study the equilibrium behaviour of a deterministic four-state
mutation-selection model as a model for the evolution of a population
of nucleotide sequences in sequence space. The mutation model is the
Kimura 3ST mutation scheme, and the selection scheme is assumed to be
invariant under permutation of sites. Considering the evolution
process both forward and backward in time, we use the ancestral
distribution as the stationary state of the backward process to derive
an expression for the mutational loss (as the difference between
ancestral and population mean fitness), and we prove a maximum
principle that determines the population mean fitness in
mutation-selection balance.
\end{abstract}

\section{Introduction}

The mathematical modelling of populations subjected to the competing
evolutionary forces of mutation and selection has a long and rich
history, see, e.g., \cite{Ewe79}.  The various approaches that have
been employed to describe DNA evolution at the molecular level can be
classified into two main categories, comprising stochastic approaches
on the one hand and deterministic on the other.

The stochastic models deal with finite populations using Wright-Fisher
sampling (\cite{Wri31,Ewe79}, Ch.\ 3). The stochastic formulation lies
at the heart of the neutral theory and had a strong influence on the
methods used today to analyse population sequence data, see, e.g.,
\cite{HC97,LG90} for reviews. Selection can be treated as well;
however, this is limited to very simple situations with two alleles
only. The analysis of more complicated settings becomes impractical.

In the deterministic formulation, more challenging selection schemes
can be treated, at the cost of neglecting genetic drift.  Classical
mutation-selection models formulate differential equations for the
evolution of gene frequencies in infinite populations
(\cite{Fis22,Hal28,CK70}, Ch.\ 6; for an up-to-date review see
\cite{Bur00}).  These have been adapted to sequence space by
identifying alleles with sequences, and choosing an appropriate
mutation mechanism --- a tradition that was started by \cite{Eig71}
and reviewed by \cite{EMS89}.  A number of results can be found, e.g.,
in \cite{OBr85,Leu86,Leu87,Rum87,Tar92} and, in a time-continuous
formulation where mutation and selection are decoupled from each
other, in \cite{Baa95,BBW97,BW01}. But this body of literature almost
exclusively deals with two-state models, where each site in the
sequence can be occupied with one of two states (or alleles), wildtype
or mutant, in the assumption that this captures the essential
behaviour of DNA sequences which are written in a four-letter
alphabet.

In a recent article, \cite{HWB01} refined this approach to a
four-state model within a physical framework, concentrating on linear
and quadratic fitness functions. The results obtained for the
four-state model show a much richer behaviour than is observed for the
two-state case.  Therefore, four-state models clearly deserve further
investigation.  In this article, we consider a four-state model for
general permutation-invariant fitness functions, formulated entirely
within the biological framework. Our main result is a maximum
principle that allows us to determine the population mean fitness in
mutation-selection balance by maximising the difference between the
fitness and a suitably defined function that describes the mutational
loss.

In practice, permutation-invariant fitness means that the fitness only
depends on the type and the number of mutations relative to a
reference sequence, not on their position in the sequence.  This is,
of course, a strong restriction; however, because of the difficult
accessibility and complexity of realistic fitness landscapes, it is a
widely used assumption in the population genetics literature (e.g.,
\cite{Cha90, Wie97}).  Furthermore, a permutation-invariant fitness
function describes the accumulation of many small mutational effects
surprisingly well. It is actually a good approximation for the fitness
function in some models for concrete experimental situations like the
DNA-binding models treated by \cite{vHB86, GH02}, which are formulated
as two-state models and would certainly profit from a generalisation
to a four-state description.

The four-state model of \cite{HWB01} is a generalisation of the
two-state model or biallelic chain \citep{BBW97}. Mutation-selection
models of this type are closely connected to certain models of
statistical physics, so-called quantum spin chains. Whereas the
biallelic chain is related to the quantum Ising chain, the four-state
model considered in this article corresponds to the Ashkin-Teller
quantum chain, see, e.g., \cite{KNK81} and \cite{Bax82}, Ch.\ 12.
However, this correspondence does not mean that results from
statistical physics can be transferred directly to biology, because
some of the quantities considered in the context of statistical
physics are not those that are of interest here, compare the
discussion by \cite{BW01}.

The outline of this paper is as follows.  We start with a general
introduction to mutation-selection models, defining the ancestral
distribution and the observable quantities in this class of models.
This is then specialised to the sequence space of the four-state model
which we are interested in, choosing appropriate models for mutation
and selection. Exploiting the permutation invariance of the fitness
function and the symmetries of the mutation model, the sequence space
can be reduced to the permutation-invariant subspace.  Subsequently,
we define the mutational loss function and state the maximum
principle.  This principle holds exactly in three special cases which
we discuss in detail. Finally, this is followed by a summary and a
brief outlook.  The rather technical proofs of the maximum principle
are given in two appendices.

\section{Mutation-selection models}

We consider a population of haploid individuals\footnote{The theory
applies as well to populations of diploids without dominance, where
the evolution equations reduce to those of a haploid population (cf.\
\cite{Bur00}, Ch. 2.2)} whose genotypes are chosen from a sequence
space, a set of a finite number $\nu$ of possible genotypes $i$. The
population is described by the population distribution
$\boldsymbol{p}$, a $\nu$-dimensional vector with entries $p_{i}\ge
0$, indicating the relative frequency of type $i$ in the population.
Hence, $\boldsymbol{p}$ has to be normalised such that
$\sum_{i=1}^{\nu}p_i=1$.  For finite population size, the $p_i$ are
rational numbers. In this article, we concentrate on the deterministic
limit of an infinite population size, where the relative frequencies
can take real values.

Ignoring environmental effects, mutation and selection are assumed to
depend only on the genotypes of individuals.  In this framework, the
evolutionary processes to be considered are birth and death of
individuals, and mutation from one type to another.

In the time-continuous model, each individual of type $i$ gives birth
to an identical copy with a rate $b_i$ and dies with a rate $d_i$,
hence we have an effective reproduction rate $r_i=b_i-d_i$, also
called the {\em Malthusian fitness}\/ (\cite{Bur00}, Ch.\ 1). These
values are collected in a diagonal {\em reproduction matrix}\/
$\boldsymbol{R}=\mbox{diag}(r_i,\ldots,r_{\nu})$.

Mutation from type $j$ to $i$ occurs with a rate $M_{ij}$. To preserve
the normalisation of the population distribution $\boldsymbol{p}$, the
diagonal entries $M_{ii}$ of the {\em mutation matrix}\/
$\boldsymbol{M}=(M_{ij})$ are chosen such that $\boldsymbol{M}$ has a
vanishing sum over the columns, $\sum_{i}M_{ij}=0$, which makes
$\boldsymbol{M}$ a Markov generator.  Unless we talk about
unidirectional mutation (cf.\ Sec.\ \ref{unidirectional mutation}), we
will assume that the mutation matrix is irreducible, i.e., each
genotype can be reached from any other by mutation, possibly in
several steps.  With the definition of the {\em time-evolution
operator}\/ $\boldsymbol{H}=\boldsymbol{R}+\boldsymbol{M}$, this leads
to the evolution equation
\begin{equation}
\label{2.1}
\boldsymbol{\dot{p}}(t)\;=\;
\big(\boldsymbol{H}-\overline{r}(t)\boldsymbol{1}\big)\,\boldsymbol{p}(t)\,,
\end{equation}
where $\overline{r}$ is the {\em population mean fitness}\/ 
$\overline{r}(t)=\sum_{i}r_{i}\,p_{i}(t)$ and $\boldsymbol{1}$ denotes the 
$\nu\times\nu$ identity matrix, cf.~\cite{CK70}, Ch.\ 6, and 
\cite{Bur00}, Ch.\ 3.

Irreducibility of $\boldsymbol{M}$ implies that of $\boldsymbol{H}$,
and the Perron-Frobenius (PF) theorem guarantees that there exists a
unique stable equilibrium solution, which is given by the strictly
positive eigenvector $\boldsymbol{p}$ corresponding to the largest
eigenvalue $\lambda_{\rm max}$ of $\boldsymbol{H}$,
\begin{equation}
\label{2.2}
\boldsymbol{H}\,\boldsymbol{p}\;=\;
\lambda_{\rm max}\,\boldsymbol{p}\, .
\end{equation}
In the limit as $t\rightarrow\infty$, the population distribution
converges towards this equilibrium solution,
$\lim_{t\rightarrow\infty}\boldsymbol{p}(t)=:\boldsymbol{p}$.

\subsection{The ancestral distribution}

We are particularly interested in the equilibrium solutions
$\dot{\boldsymbol{p}}=0$. In this case, Eq.~(\ref{2.1}) becomes an
eigenvalue equation for $\boldsymbol{H}$ with eigenvalue $\lambda_{\rm
max}=\overline{r}$.  The {\em right}\/ PF eigenvector $\boldsymbol{p}$
is the population distribution in equilibrium, whereas the entries
$z_i$ of the {\em left}\/ PF eigenvector $\boldsymbol{z}$ determine
the relative reproductive success of type-$i$ individuals, as shown by
\cite{HRWB02}. The {\em ancestral distribution}, also introduced by
\cite{HRWB02}, is a probability distribution defined as
$a_{i}=z_{i}p_{i}$, with the normalisation of $\boldsymbol{z}$ chosen
such that $\sum_{i} a_{i}=1$. Here, $a_{i}$ specifies the fraction of
the equilibrium population whose ancestors, an infinitely long time
ago, were of type $i$.

In analogy to the way that the population distribution is defined as a
time-dependent quantity, this can also be done for the relative
reproductive success $\boldsymbol{z}$ and the ancestral distribution
$\boldsymbol{a}$. However, this demands some notational efforts, and
as we do not need this property later on, we limit ourselves to the
definition of the ancestral distribution as an equilibrium quantity.

\subsection{Means}

A population is macroscopically described by mean quantities. We
introduced two probability distributions, hence there are two types of
averages that are relevant in our model. Every mapping $o$ that
assigns a value $o_i$ to each possible genotype $i$ can be averaged
with respect to the population distribution or the ancestral
distribution.

The {\em population mean}\/ of $o$, denoted by $\overline{o}(t)$, is
given by
\begin{equation}
\label{2.3}
\overline{o}(t)\; :=\; \sum_{i} o_{i}\,  p_{i}(t)\, .
\end{equation}
The population mean in equilibrium, i.e., in the limit as
$t\rightarrow\infty$, is denoted by $\overline{o}$.

The {\em ancestral mean}\/ of an operator $o$, denoted by
$\widehat{o}$, is defined as
\begin{equation}
\label{2.4}
\widehat{o}\; :=\; \sum_{i} o_{i}\, a_{i}\,.
\end{equation}
Note that the ancestral mean does not depend on time, as we defined
the ancestral distribution as an equilibrium quantity only.

\section{The four-state model}

The genetic information is coded in the DNA as a string composed of
the purines adenine and guanine ($A$,$G$) and the pyrimidines cytosine
and thymine ($C$,$T$).

We consider DNA strands of fixed length $N$, which may, for instance,
code for an enzyme, as modelled by \cite{HWB01}.  The four basic
states $\{A,G,C,T\}$ are mapped onto $\{0,1,2,3\}$, or, as it is done
by \cite{HWB01}, onto $\{++, +-, -+, --\}$. Conveniently, one can
exploit the freedom in the choice of this mapping by introducing a
relative rather than an absolute mapping between the bases
$\{A,G,C,T\}$ and the symbols $\{0,1,2,3\}$. This essentially means
that one can choose independent mappings at each position along the
DNA strand.  At any position, the mapping can be defined such that the
symbol $0$, or $++$, corresponds to the corresponding base in a given
preferred sequence, which is usually chosen to be the {\em wildtype}\/
or {\em master sequence}\/ of maximal fitness $r_{\rm max}$.  Thus the
wildtype sequence is mapped onto the sequence $\{0\}^N=000\ldots 0$,
see \cite{HWB01} for details. The mapping between the remaining
nucleotides and the symbols $1,2,3$ will be discussed below.  The
sequence space consists of all possible $N$-letter sequences in these
four symbols, so it is given by $\{0,1,2,3\}^N$ and has dimension
$4^N$.

In what follows, we shall not really need the complete information
about the sequences. It will be sufficient to characterise a sequence
by its {\em mutational distance}\/ with respect to the wildtype
sequence $\{0\}^N$, which just counts the deviations from the wildtype
sequence.  Whereas in the two-state model the mutational distance is
given by a single integer, which counts the number of bases along the
DNA strand that differ from those in the wildtype, we now need three
non-negative integers $d_1$, $d_2$ and $d_3$, according to the
different types of mutations that can occur.  We define the mutational
distance $\boldsymbol{d}$ of a sequence as
\begin{equation}
\label{3.1}
\boldsymbol{d}\; =\; 
\begin{pmatrix}d_1\cr d_2 \cr d_3\end{pmatrix}\;:=\;
\begin{pmatrix}\#(1)\cr \#(2)\cr \#(3)\end{pmatrix}\, ,
\end{equation}
where $\#(1)$, $\#(2)$ and $\#(3)$ denote the number of entries $1$,
$2$ and $3$ in the sequence, respectively. The total mutational
distance is defined as the sum $d:=d_1+d_2+d_3$, which takes values
$0\le d\le N$.

\subsection{Mutation}

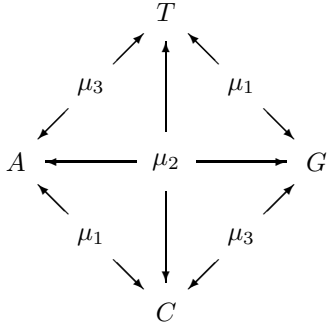
\begin{figure}[htb]
\begin{center}
\setlength{\unitlength}{1cm}
\begin{picture}(6,4.5)(0,0.75)
\put(1,3){\makebox(0,0){$A$}}
\put(3,1){\makebox(0,0){$C$}}
\put(3,5){\makebox(0,0){$T$}}
\put(5,3){\makebox(0,0){$G$}}
\multiput(2,2)(2,2){2}{\makebox(0,0){$\mu_1$}}
\put(3,3){\makebox(0,0){$\mu_2$}}
\multiput(4,2)(-2,2){2}{\makebox(0,0){$\mu_3$}}
\put(3.4,3){\vector(1,0){1.2}}
\put(2.6,3){\vector(-1,0){1.2}}
\put(3,3.4){\vector(0,1){1,2}}
\put(3,2.6){\vector(0,-1){1.2}}
\multiput(2.3,4.3)(2,-2){2}{\vector(1,1){0.4}}
\multiput(1.7,3.7)(2,-2){2}{\vector(-1,-1){0.4}}
\multiput(1.7,2.3)(2,2){2}{\vector(-1,1){0.4}}
\multiput(2.3,1.7)(2,2){2}{\vector(1,-1){0.4}}
\end{picture}
\caption{The Kimura 3 ST mutation scheme.}
\label{fig1}
\end{center}
\end{figure}

Mutation is taken to be a point process that acts at each site
independently. Disregarding more complicated mechanisms such as
deletions and insertions, we only take into account the replacement of
one base by another. Taken over the whole sequence, this happens with
certain rates $\mu_k$, where $k$ indicates the type of replacement.
We allow only one mutation at a time, as modelled by a Poisson
process. This leads to a single step mutation model, which was first
introduced by \cite{OK73}.  We work with the {\em Kimura 3ST mutation
scheme}\/ shown in Fig.~\ref{fig1} (\cite{Kim81, SOWH96, EG01}, Ch.\
13) which assumes that, of a possible 12 mutation rates that can be
chosen in our setting, only three different mutation rates $\mu_{1}$,
$\mu_{2}$ and $\mu_{3}$ occur. In particular, forward and backward
mutation rates are the same, and the mutation process respects a
symmetry between exchanges of purines and pyrimidines.

This mutation scheme can be treated to various degrees of
sophistication. Apart from the full Kimura 3ST scheme, where all three
mutation rates are different, there are also two simpler models that
are worth mentioning.

The simplest approach is to take all mutation rates to be equal,
$\mu_1=\mu_2=\mu_3$. This case is known as the {\em Jukes-Cantor
mutation scheme} \citep{JC69}.

Due to the similar shapes of the nucleotides, {\it transitions}, i.e.,
the replacement of one purin/ pyrimidine by the other, are more
frequent than {\it transversions}, i.e., the replacement of a purin/
pyrimidine by a pyrimidine/purin. The mutation rates describing the
transversions are fairly similar, so $\mu_1\approx\mu_3$, whereas the
mutation rate for the transitions $\mu_2$ is typically larger by a
factor of about $2$ to $40$.  This is taken into account in the {\em
Kimura 2 parameter model}\/ \citep{Kim80} by assuming that
$\mu_1=\mu_3<\mu_2$.

Assume there is one particular sequence $s_0$ with maximal fitness
$r_{\rm max}$, the wildtype or master sequence, which is mapped onto
$\{0\}^N$.  For any other sequence, the corresponding representation
in terms of the symbols $\{0,1,2,3\}$ is then obtained by comparing it
to the wildtype, and assigning one of the labels $1$, $2$, $3$ at each
position where it differs from the wildtype sequence, according to the
type of mutation as given in Fig.~\ref{fig1}. Analogously to the
mutational distance, we can define the {\em Hamming distance}\/
(\cite{Ham50,vLi82}, Ch.\ 3) between two sequences $s_i$ and $s_j$, by
comparing the sequences with each other.  The restricted Hamming
distances $d_k(s_i,s_j)$ are the numbers of type-$k$ mutations between
the sequences $s_{i}$ and $s_{j}$, i.e., mutations that occur with
rate $\mu_k$; the total Hamming distance is
$d(s_i,s_j)=d_1(s_i,s_j)+d_2(s_i,s_j)+d_3(s_i,s_j)$.

In the Kimura 3ST setting, the entries $M_{ij}$ of the mutation matrix
are given by
\begin{equation}
\label{3.2}
M_{ij}=\begin{cases}
{\displaystyle\frac{\mu_k}{N}} & 
\mbox{for $d(s_i,s_j)=d_k(s_i,s_j)=1\,$,} \\[1ex] 
0 & \mbox{for $d(s_i,s_j)>1\,$,} \\ 
{\displaystyle -\sum_{\ell\ne i}M_{\ell i}=-\sum_{k=1}^3\mu_k} & 
\mbox{for $i=j\,$,} 
\end{cases}
\end{equation}
where, as mentioned above, the diagonal entries are chosen such that
$\boldsymbol{M}$ is a Markov generator.  Here, the mutation rates are
scaled as mutation rates per site, with the mutation rate over the
whole DNA string being constant, see the discussion in \cite{BW01}.

\subsection{Selection}

Whereas the process of mutation is well understood and straightforward
to model, the choice of the fitness landscape on the molecular level
is far from being clear.  Realistic fitness landscapes would be rather
rugged and strongly dependent on the function of the DNA sequence, but
they are hard to access experimentally.

We shall use the severe simplification of a permutation-invariant
fitness function, which is nevertheless a rather common (and usually
implicitly made) assumption in theoretical investigations of
mutation-selection models such as \cite{Cha90, Wie97} as well as in
the modelling of concrete experimental settings like in DNA-binding
models \citep{vHB86, GH02}.  Using permutation-invariant fitness, one
assumes that the fitness of a sequence depends only on the {\em
number}\/ of mutations of the various kinds, {\em not}\/ on their
location within the sequence. Hence, we can describe a sequence
completely by its mutational distance
$\boldsymbol{d}=(d_{1},d_{2},d_{3})$ with respect to the wildtype
sequence.  As there are $4^N$ different sequences, but only
$(N+1)(N+2)(N+3)/6$ different distances $\boldsymbol{d}$ with $0\le d
\le N$, this reduces the effective type space enormously. In the
permutation-invariant fitness model, the number of possible different
genotypes, and thus the dimension of the permutation-invariant
subspace, is given by $\nu=(N+1)(N+2)(N+3)/6$.

\section{Reduction to the permutation-invariant subspace}

In our model, three different spaces are relevant: (i) the
$4^N$-dimensional full sequence space, (ii) the reduced sequence space
of dimension $\nu$, and (iii) the three-dimensional space of the
mutational distances.

In the {\em full sequence space}\/ of dimension $4^N$, each sequence
corresponds to a different basis vector, and the population
$\boldsymbol{p}$ is then completely determined as a point on the
$(4^N-1)$-dimensional hyperplane defined by $\sum_{i=1}^{4^N}p_i=1$,
where the projection on each axis gives the frequency $p_i$ of the
corresponding sequence.

Analogously, the $\nu$-dimensional {\em reduced sequence space}, which
is the permutation-invariant subspace of the full sequence space, is
spanned by unit vectors, each of which corresponds to the set of all
sequences which have the same number of mutations of each type, i.e.,
to one of the $\nu$ different mutational distances
$\boldsymbol{d}$. Here, the population $\boldsymbol{p}$ is also given
as a point on a hyperplane (of dimension $\nu-1$). In general, the
transition from the full to the reduced sequence space is accompanied
by a loss of information.  As long as we consider systems with a
unique equilibrium population, we know that this equilibrium will be
permutation-invariant, because starting from a permutation-invariant
initial population, we will reach a permutation-invariant equilibrium
because the fitness function, as well as the mutation scheme,
disregard the order in the sequences. As the equilibrium is unique, it
will be reached from any initial condition. Therefore, sequences with
the same numbers of each type of mutation, i.e., the same
$\boldsymbol{d}$, must occur with the same frequency in the
equilibrium population. Thus, as long as we are only interested in
equilibrium properties of systems with a unique equilibrium, it
suffices to restrict ourselves to the reduced sequence space.

\begin{figure}[hbt]
  \begin{center}
  \includegraphics[width=6cm]{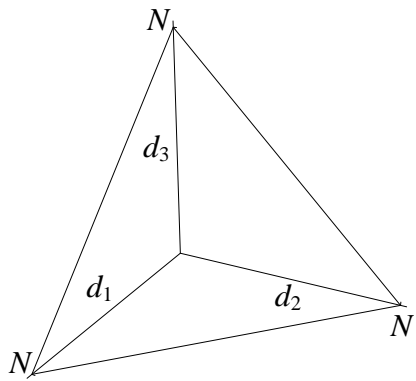}
  \caption{Mutational distance space in the case of permutation-invariant 
fitness.}  
  \label{fig2}
  \end{center}
\end{figure}

Finally, we have the three-dimension {\em mutational distance space}\/
of the mutational distance vectors $\boldsymbol{d}$ with Cartesian
coordinates $d_{1}$, $d_{2}$ and $d_{3}$. The basis of this space is
formed by the Cartesian unit vectors $\boldsymbol{e}_1=(1,0,0)^t$,
$\boldsymbol{e}_2=(0,1,0)^t$ and $\boldsymbol{e}_3=(0,0,1)^t$, which
are the basic directions of mutation. The condition $0\le
d=d_{1}+d_{2}+d_{3}\le N$ restricts the possible mutational distance
vectors $\boldsymbol{d}$ to a simplex in the positive quadrant, as
shown in Fig.~\ref{fig2}. There is a one-to-one correspondence between
the sequences in the reduced sequence space and the mutational
distance vectors $\boldsymbol{d}$ in this simplex, and we label the
elements of the reduced sequence space by the corresponding mutational
distance vectors $\boldsymbol{d}$. Whenever we speak of a sequence
$\boldsymbol{d}$, we refer to the corresponding sequences in the
reduced sequence space.

The mutational distance space is required to define the neighbourhood
of sequences. As we use a single step mutation model, each sequence,
now labelled by $\boldsymbol{d}$, has at most 12 neighbours, i.e.,
sequences to which they can mutate within a single mutation step.  The
neighbours of a sequence $\boldsymbol{d}$ have mutational distances
$\boldsymbol{d}\pm\boldsymbol{e}_{\xi}$, where $\xi$ determines the
direction of mutation and the $\boldsymbol{e}_{\xi}$ are combinations
of the basis vectors $\boldsymbol{e}_k$. More precisely, there are two
types of mutational directions $\xi$; firstly, the mutations from
wildtype to mutant, $\xi=k$ with $k\in\{1,2,3\}$, which correspond to
unit vectors $\boldsymbol{e}_{\xi}=\boldsymbol{e}_k$, and vice versa,
which correspond to $\boldsymbol{e}_{\xi}=-\boldsymbol{e}_k$.  For the
second type, where one type of mutation is replaced by another
mutation, one mutation step corresponds to a vector
$\boldsymbol{e}_{\xi}=\boldsymbol{e}_k-\boldsymbol{e}_{\ell}$. These
directions are labelled by pairs $\xi=(+k,-\ell)$ with
$k,\ell\in\{1,2,3\}$ and $k>\ell$, and the corresponding mutations in
the inverse directions are labelled accordingly with $k<\ell$.
Finally, we note that points on the surface of the simplex in
mutational distance space have fewer neighbours. Clearly, only those
mutations that do not leave the simplex are permitted.

\subsection{Similarity transformation}

We exploit the permutation invariance of the fitness function to
reduce the sequence space to its relevant permutation-invariant
subspace.  Therefore, we transform our time-evolution operator
$\boldsymbol{H}$, which is a $4^N\times4^N$-matrix, to a matrix
$\boldsymbol{H}_{\rm piv}$ of dimension $\nu=(N+1)(N+2)(N+3)/6$ that
describes the reduced sequence space only, the subscript ``piv''
refers to the {\bf p}ermutation-{\bf i}n{\bf v}ariant subspace.  This
is done by the means of a similarity transformation
$\boldsymbol{T}$. We have
\begin{equation}
\label{4.1}
\boldsymbol{T}^{-1}\boldsymbol{HT}\;=\;
\begin{pmatrix}\boldsymbol{H}_{\rm piv} & (*)\cr
 0 & \boldsymbol{H'}\end{pmatrix}\;,
\end{equation}
where $\boldsymbol{H}_{\rm piv}$ is the $\nu$-dimensional
time-evolution operator describing the reduced sequence space.

A condition on the transformation $\boldsymbol{T}$ is that it
preserves the Markov property of the mutation matrix
$\boldsymbol{M}$. Hence, it must be an $L^1$-transformation, which is
guaranteed by the property $\sum_{i}T_{ij}=1$.  For the relevant
permutation-invariant subspace, we need to combine all sequences that
belong to the same mutational distance vector $\boldsymbol{d}$. This,
together with the Markov condition, determines the entries in the
first $\nu$ columns of $\boldsymbol{T}$.  In the column assigned to
$\boldsymbol{d}$, each entry is either $0$, or, if the sequence
corresponding to the row in question has the mutational distance
$\boldsymbol{d}$, it is $1/n_{\boldsymbol{d}}$, where
$n_{\boldsymbol{d}}$ is the number of sequences that are mapped onto
$\boldsymbol{d}$. This number is given by the multinomial coefficients
\begin{equation}
\label{4.2}
n_{\boldsymbol{d}} \;=\;
\begin{pmatrix} N\cr
d_0,d_1,d_2,d_3\end{pmatrix}\;=\;\frac{N!}{d_0!\,d_1!\,d_2!\,d_3!}\, ,
\end{equation}
with $d_0:=N-\sum_{k=1}^3d_k\,$ denoting the number of wildtype sites.
The actual choice of other columns of $\boldsymbol{T}$ does not
influence the submatrix $\boldsymbol{H}_{\rm piv}$, here we only need
that $\boldsymbol{T}$ is invertible.

For a sequence length of $N=2$, for example, the relevant part of the 
transformation $\boldsymbol{T}$ has the form
\begin{equation}
\label{4.2i}
\boldsymbol{T}=
\begin{array}{c@{\!\!\!\!}c@{}c@{}c@{}c@{}c@{}c@{}c@{}c@{}c@{}c@{\;\;}c}
&000 & 100 & 010 & 001 & 200 & 110 & 101 & 020 & 011 & 002 & \\[2ex]
\left(
\begin{array}{c}
   \\   \\   \\   \\  \\   \\   \\   \\   \\   \\   \\   \\   \\   \\  \\  
\vphantom{0}\end{array}\right. &
\begin{array}{c} 
 1  \\   \\   \\   \\   \\   \\   \\   \\   \\   \\ \\   \\   \\   \\   \\  
 \hphantom{1/2}\vphantom{0}\end{array}& \begin{array}{c}
 \\ 1/2\\   \\   \\ 1/2\\   \\   \\   \\   \\   \\   \\   \\   \\   \\   \\ 
 \vphantom{0}\end{array}& \begin{array}{c}
 \\   \\ 1/2\\   \\   \\   \\   \\   \\ 1/2\\   \\   \\   \\   \\   \\   \\ 
 \vphantom{0}\end{array}& \begin{array}{c}
 \\   \\   \\ 1/2\\   \\   \\   \\   \\   \\   \\   \\   \\ 1/2\\   \\   \\ 
 \vphantom{0}\end{array}& \begin{array}{c} 
 \\   \\   \\   \\   \\  1 \\   \\   \\   \\   \\   \\   \\   \\   \\   \\ 
 \hphantom{1/2}\vphantom{0}\end{array}& \begin{array}{c}
 \\   \\   \\   \\   \\   \\ 1/2\\   \\   \\ 1/2\\   \\   \\   \\   \\   \\ 
 \vphantom{0}\end{array}& \begin{array}{c}
 \\   \\   \\   \\   \\   \\   \\ 1/2\\   \\   \\   \\   \\   \\ 1/2\\   \\ 
 \vphantom{0}\end{array}& \begin{array}{c}
 \\   \\   \\   \\   \\   \\   \\   \\   \\   \\ 1 \\   \\   \\   \\   \\ 
 \hphantom{1/2}\vphantom{0}\end{array}& \begin{array}{c}
 \\   \\   \\   \\   \\   \\   \\   \\   \\   \\   \\ 1/2\\   \\   \\ 1/2\\ 
 \vphantom{0}\end{array}& \begin{array}{c}
 \hphantom{1/2}
 \\   \\   \\   \\   \\   \\   \\   \\   \\   \\   \\   \\   \\   \\   \\  1 
\vphantom{0}\end{array}& 
\left|\;\;
\begin{array}{c}  
\\ \\ \\ \\ \\ \\ \\  (*) \\ \\ \\ \\ \\ \\ \\ \\ 
 \vphantom{0}\end{array}
\;\right)\;
\begin{array}{c}
00\\01\\02\\03\\10\\11\\12\\13\\20\\21\\22\\23\\30\\31\\32\\33
\vphantom{0}\end{array}
\end{array}
\end{equation}
where the triples at the top give the mutational distances
$\boldsymbol{d}$ to which each column corresponds, whereas on the
right the actual sequences $s$ corresponding to each line are
displayed. Only non-zero entries are shown. In this case, the
remaining six columns of $\boldsymbol{T}$, shown symbolically as
$(*)$, correspond to the antisymmetric subspace, which contains
sequences with mutational distances
$\boldsymbol{d}=(100)^t,(010)^t,(001)^t,(110)^t,(101)^t,(011)^t$.

The diagonal entries of $\boldsymbol{H}_{\rm piv}$ remain unchanged
compared to the original $\boldsymbol{H}$; they are
$\boldsymbol{H}_{\rm piv,\boldsymbol{d}\boldsymbol{d}}=
r_{\boldsymbol{d}}-\sum_k\mu_k$.  The off-diagonal entries, i.e., the
mutation rates $u$ in the permutation-invariant subspace, depend on
the direction of mutation. Using the normalised versions of the
mutational distances $x_k:=d_k/N$, they are given by
\begin{equation}
\label{4.3}
\begin{array}{l@{\qquad}l@{\;=\;}r@{\qquad}l}
\boldsymbol{d}\rightarrow \boldsymbol{d}+\boldsymbol{e}_k:\; & 
u^{+k}_{\boldsymbol{d}}&\mu_{k}\,x_{0} & \mbox{(3 eqns.)} \\
\boldsymbol{d}\rightarrow \boldsymbol{d}-\boldsymbol{e}_k:\; & 
u^{-k}_{\boldsymbol{d}}&\mu_{k}\,x_{k} & \mbox{(3 eqns.)} \\
\boldsymbol{d}\rightarrow \boldsymbol{d}+\boldsymbol{e}_k-
\boldsymbol{e}_{\ell}:\; & 
u^{+k,-\ell}_{\boldsymbol{d}}&\mu_{m}\,x_{\ell} & \mbox{(6 eqns.)} 
\end{array}
\end{equation}
where $k,l,m\in \{1,2,3\}$ are pairwise different, so
$\{k,l,m\}=\{1,2,3\}$. Our notation is such that
$u^{+k}_{\boldsymbol{d}}$ and $u^{-k}_{\boldsymbol{d}}$ denote the
rates for mutations from distance $\boldsymbol{d}$ in the positive and
negative $k$ direction, respectively, and
$u^{+k,-\ell}_{\boldsymbol{d}}$ the corresponding rate in direction
$\boldsymbol{e}_{k}-\boldsymbol{e}_{\ell}$.  The mutation rates now
depend on $\boldsymbol{d}$, reflecting the fraction of sites that can
mutate with the specified effect.  In particular, note that this
implies that $\boldsymbol{H}_{\rm piv}$ is not a symmetric matrix.
The above definition of mutation rates takes care of the boundary
condition $u^{\pm\xi}_{\boldsymbol{d}}=0$ for $\boldsymbol{d}$ on the
boundary of the relevant simplex in the mutational distance space with
$\pm\boldsymbol{e}_{\xi}$ pointing outwards.

\subsection{Symmetrisation of the time-evolution operator 
$\boldsymbol{H}_{{\bf\rm piv}}$}

There is an alternative way to arrive at a matrix that describes the
permutation-invariant subspace, which, however, does not preserve the
Markov property for $\boldsymbol{M}$. As the original $4^N\times4^N$
matrix $\boldsymbol{H}$ is real symmetric, we can block-diagonalise it
by an orthogonal transformation $\boldsymbol{O}$.  This is an
$L^2$-transformation, and it preserves the symmetry of the matrix, so
the corresponding $\nu\times \nu$ matrix
$\widetilde{\boldsymbol{H}}_{\rm piv}$ given by
\begin{equation}
\label{4.4}
\boldsymbol{O}^{t}\boldsymbol{H}\boldsymbol{O}\;=\;
\begin{pmatrix}\widetilde{\boldsymbol{H}}_{\rm piv} & 0 \cr
0 & \widetilde{\boldsymbol{H}}'\end{pmatrix}
\end{equation}
is symmetric, where $\boldsymbol{O}^{t}=\boldsymbol{O}^{-1}$ denotes
the transpose of the orthogonal matrix $\boldsymbol{O}$.  In this
case, the states in the permutation-invariant subspace are again
superpositions of all sequences of equal distance $\boldsymbol{d}$,
but now with coefficients $1/\sqrt{n_{\boldsymbol{d}}}$, as opposed to
$1/n_{\boldsymbol{d}}$ for the $L^1$-transformation $\boldsymbol{T}$.
Knowing this connection, we can symmetrise the permutation-invariant
part $\boldsymbol{H}_{\rm piv}$ of the time-evolution operator. By
slight abuse of notation, we get for the permutation-invariant part
\begin{equation}
\label{4.5}
\widetilde{\boldsymbol{H}}_{\rm piv}\;=\;
\big(\boldsymbol{O}^t\boldsymbol{H}\boldsymbol{O}\big)_{\rm piv}\;=\;
\big(\boldsymbol{O}^t\boldsymbol{T}\boldsymbol{T}^{-1}\boldsymbol{H}
\boldsymbol{T}\boldsymbol{T}^{-1}\boldsymbol{O}\big)_{\rm piv}
\;=\;\boldsymbol{D}^{-1}
\boldsymbol{H}_{\rm piv}\boldsymbol{D}\, ,
\end{equation}
where $\boldsymbol{D}:=(\boldsymbol{T}^{-1}\boldsymbol{O})_{\rm piv}$,
and the subscripts mean that we restrict to the permutation-invariant
part.  As the relevant columns of the matrices $\boldsymbol{T}$ and
$\boldsymbol{O}$ differ only by factors $\sqrt{n_{\boldsymbol{d}}}$,
the transformation $\boldsymbol{D}$ that symmetrises the
time-evolution operator $\boldsymbol{H}_{\rm piv}$ is in fact {\em
diagonal}, with entries
$D_{\boldsymbol{d}\boldsymbol{d}}=\sqrt{n_{\boldsymbol{d}}}$.

The corresponding mutation rates $\widetilde{u}$ of the symmetrised
system are given by
\begin{eqnarray}
\widetilde{u}^{+k}_{\boldsymbol{d}}&=&
\sqrt{\frac{n_{\boldsymbol{d}}}{n_{\boldsymbol{d}+\boldsymbol{e}_k}}}
u^{+k}_{\boldsymbol{d}}
\;\:=\;\:\mu_k\sqrt{x_0(x_k+\frac{1}{N})}
\;\:=\;\:
\sqrt{u^{+k}_{\boldsymbol{d}}u^{-k}_{\boldsymbol{d}+\boldsymbol{e}_k}}\, ,
\nonumber\\
\widetilde{u}^{-k}_{\boldsymbol{d}}&=&
\sqrt{\frac{n_{\boldsymbol{d}}}{n_{\boldsymbol{d}-\boldsymbol{e}_k}}}
u^{-k}_{\boldsymbol{d}}
\;\:=\;\:\mu_k\sqrt{x_k(x_0+\frac{1}{N})}
\;\:=\;\:
\sqrt{u^{-k}_{\boldsymbol{d}}u^{+k}_{\boldsymbol{d}-\boldsymbol{e}_k}}\, ,
\nonumber\\
\widetilde{u}^{+k,-\ell}_{\boldsymbol{d}}
&=&
\sqrt{\frac{n_{\boldsymbol{d}}}{n_{\boldsymbol{d}+\boldsymbol{e}_k-
\boldsymbol{e}_{\ell}}}}u^{+k,-\ell}_{\boldsymbol{d}}
\;\:=\;\:\mu_m\sqrt{x_{\ell}(x_k+\frac{1}{N})} 
\;\:=\;\:\sqrt{u^{+k,-{\ell}}_{\boldsymbol{d}}
u^{+\ell,-k}_{\boldsymbol{d}+\boldsymbol{e}_k-\boldsymbol{e}_{\ell}}}\, .
\label{4.6}
\end{eqnarray}

The property that $\boldsymbol{H}_{\rm piv}$ is symmetrisable by means
of a diagonal transformation allows us to write the eigenvalue
equation (\ref{2.2}) in ancestral formulation, which is the starting
point for the proofs of the maximum principle. In fact, the proofs
presented below can be formulated for {\em any}\/ mutation matrix that
is symmetrisable by a diagonal transformation, which is equivalent to
the property that it describes a reversible process,
\begin{equation}
\label{4.6a}
M_{ij}\,q_j\;=\;M_{ji}\,q_i\, ,
\end{equation}
where $\boldsymbol{q}$ is the equilibrium distribution of the mutation
process without selection.  In this case, we can use the diagonal
transformation $Q_{ij}=\delta_{ij}\sqrt{q_j}$, and from (\ref{4.6a}),
we obtain the symmetrised mutation rates
$\widetilde{M}_{ij}=M_{ij}\sqrt{q_j/q_i}=\sqrt{M_{ij}M_{ji}}$.
Therefore, the mutational distances are not bound to be the numbers of
mutations of a DNA sequence, but the model can be reinterpreted in the
multilocus model context, where the genetic distances are three
arbitrary traits that determine the fitness. In fact, the maximum
principle discussed below can also be derived for a model with $n$
states at each site of the sequence, which then has an interpretation
as $n$ traits contributing to the fitness, as long as we talk about a
$n$-dimensional single step model with a mutation matrix that
describes a reversible process. A more general and more systematic
approach thus seems feasible, it will be described by \cite{BBBK}.

In what follows, we drop the subscript, and use $\boldsymbol{H}$ and
$\widetilde{\boldsymbol{H}}$ to denote the submatrices corresponding
to the permutation-invariant subspace.  Both time-evolution operators
$\boldsymbol{H}$ and $\widetilde{\boldsymbol{H}}$ have the same
eigenvalues, but the eigenvectors differ. The left and right
eigenvectors $\widetilde{\boldsymbol{z}}$ and
$\widetilde{\boldsymbol{p}}$ of $\widetilde{\boldsymbol{H}}$, for the
largest eigenvalue, are related to the corresponding eigenvectors
$\boldsymbol{z}$ and $\boldsymbol{p}$ of $\boldsymbol{H}$ by
\begin{equation}
\label{4.7}
\widetilde{\boldsymbol{z}}\;=\;
\boldsymbol{z}\boldsymbol{D} \quad\mbox{and}\quad 
\widetilde{\boldsymbol{p}}\;=\;
\boldsymbol{D}^{-1}\boldsymbol{p}\;.
\end{equation}
The relation between the ancestral distribution $\boldsymbol{a}$ and
the symmetrised population $\widetilde{\boldsymbol{p}}$ is given by
\begin{equation}
\label{4.8}
a_i\;=\;z_i\,p_i\;=\;
(\widetilde{\boldsymbol{z}}\boldsymbol{D}^{-1})_i\,
(\boldsymbol{D}\widetilde{\boldsymbol{p}})_i
\;=\;\widetilde{z}_i\,\widetilde{p}_i\;\sim\;{\widetilde{p}_{i}}^{\,2} \;,
\end{equation}
as $\widetilde{\boldsymbol{z}}\sim\widetilde{\boldsymbol{p}}$ due to
the symmetry of $\widetilde{\boldsymbol{H}}$.  With the relation
$\widetilde{\boldsymbol{p}}\sim\sqrt{\boldsymbol{a}}$, the eigenvalue
equation of $\widetilde{\boldsymbol{H}}$ in ancestral formulation
becomes $\widetilde{\boldsymbol{H}}\sqrt{\boldsymbol{a}}=
\overline{r}\sqrt{\boldsymbol{a}}$, which explicitly reads
\begin{equation}
\label{4.9}
\Big[r_{\boldsymbol{d}}-\sum_{\xi}\left(u^{+\xi}_{\boldsymbol{d}}+
u^{-\xi}_{\boldsymbol{d}}\right)\Big]\sqrt{a_{\boldsymbol{d}}}
\;+\;
\sum_{\xi}\left[
\widetilde{u}^{+\xi}_{\boldsymbol{d}-\boldsymbol{e}_{\xi}}
\sqrt{a_{\boldsymbol{d}-\boldsymbol{e}_{\xi}}}
+\widetilde{u}^{-\xi}_{\boldsymbol{d}+\boldsymbol{e}_{\xi}}
\sqrt{a_{\boldsymbol{d}+\boldsymbol{e}_{\xi}}}
\right] 
\;=\;
\overline{r}\sqrt{a_{\boldsymbol{d}}} 
\end{equation}
for some distance $\boldsymbol{d}$.  Here, $\xi$ determines the six
possible directions of mutation, and the sign indicates the forward
and backward direction.

\section{Maximum principle}

For the permutation-invariant system, we can derive a maximum
principle for the population mean fitness that involves maximisation
only over the three components of $\boldsymbol{x}=\boldsymbol{d}/N$ of
the mutational distance space, as opposed to the maximisation over the
$\nu$-dimensional reduced sequence space according to Rayleigh's
principle. It finds its analogue in the scalar maximum principle given
in \cite[Eqs.~(30) and (33)]{HRWB02}.  In the four-state model, we
have
\begin{equation}
\label{5.1}
\overline{r}\;=\;
\sup_{\boldsymbol{x}}\big(r(\boldsymbol{x})-g(\boldsymbol{x})\big)
\end{equation}
with the mutational loss function $g(\boldsymbol{x})$ defined as
\begin{equation}
\label{5.2}
g(\boldsymbol{x})\;:=\;
\sum_{\xi}\Bigg(u^{+\xi}(\boldsymbol{x})+
u^{-\xi}(\boldsymbol{x})-
2\sqrt{u^{+\xi}(\boldsymbol{x})u^{-\xi}(\boldsymbol{x})}\,\Bigg)
\end{equation}
with summation over all six directions of mutation $\xi$.

This is exact for three special cases, namely (i) for unidirectional
mutation, (ii) for linear fitness and mutation functions, and, most
importantly, (iii) in the limit of infinite sequence length. These
cases will be explained in what follows, and the derivation of the
maximum principle for each case is presented as well.  For other
systems, the maximum principle gives an approximation, which, under
reasonable assumptions, one might expect to differ from the true
result by correction terms of order $1/N$.

If the supremum in Eq.~(\ref{5.1}) is assumed at a unique value
$\boldsymbol{x}$, which is the generic case, this value is the
ancestral mean mutational distance $\widehat{\boldsymbol{x}}$, and we
have, in addition to Eq.~(\ref{5.1}),
\begin{equation}
\label{5.3}
\overline{r}\;=\;
r(\widehat{\boldsymbol{x}})-g(\widehat{\boldsymbol{x}})
\;=\;\widehat{r}-g(\widehat{\boldsymbol{x}}) \, ,
\end{equation}
which again is exact for the three special cases mentioned
before. This maximum principle, in the terminology of physics, is akin
to the principle of minimal free energy.

\subsection{Unidirectional mutation} \label{unidirectional mutation}

We now consider the first of the three situations where the maximum
principle is exact, the case of unidirectional mutation.

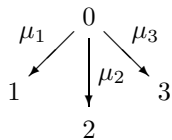
\begin{figure}[hbt]
\begin{center}
\setlength{\unitlength}{1cm}
\begin{picture}(2,2)
\put(0,0.5){\makebox(0,0){1}}
\put(1,0){\makebox(0,0){2}}
\put(2,0.5){\makebox(0,0){3}}
\put(1,1.5){\makebox(0,0){0}}
\put(0.8,1.3){\vector(-1,-1){0.6}}
\put(1,1.2){\vector(0,-1){0.9}}
\put(1.2,1.3){\vector(1,-1){0.6}}
\put(0.25,1.25){\makebox(0,0){$\mu_1$}}
\put(1.3,0.7){\makebox(0,0){$\mu_2$}}
\put(1.75,1.25){\makebox(0,0){$\mu_3$}}
\end{picture}
\caption{Simplified mutation scheme for unidirectional mutation.}
\label{fig3}
\end{center}
\end{figure}

Unidirectional mutation means that only such mutations that increase
$d$ happen, the mutation rates towards types with smaller or equal $d$
are zero.  In the most important case of a monotonically decreasing
fitness function, this means that all mutations are deleterious. This
mutation scheme does not go in line with the mutation rates for the
DNA system as given in Eq.~(\ref{4.3}), but it is equivalent with a
collapse of the mutation scheme as shown in Fig.~\ref{fig3}.

\begin{figure}[hbt]
  \begin{center}
  \includegraphics[width=7cm]{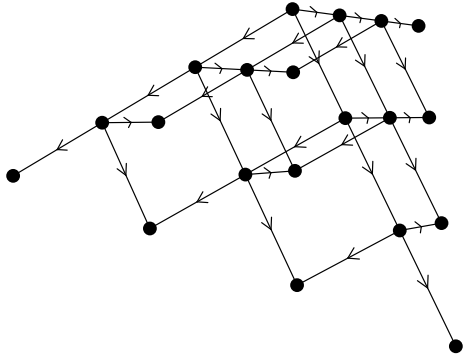}
  \caption{Structure of the mutational distance space for 
unidirectional mutation.}  
  \label{fig4}
  \end{center}
\end{figure}

For unidirectional mutation, the mutation matrix $\boldsymbol{M}$ is
no longer irreducible. Hence, in this case, the Perron-Frobenius
theorem does not apply. The equilibrium is not unique, but depends on
initial conditions. Once the wildtype is lost in the population, it
can never occur again, because the mutation rates back to types with
smaller $d$ are zero. The structure of the mutational distance space
is shown in Fig.~\ref{fig4}, where the wildtype on the top corner of
the mutational distance space ``feeds'' all mutants underneath.

Although unidirectional mutation rates do not represent the mutation
model for DNA sequences as set up in this article, they still are a
reasonable approximation. If in the DNA model the selection is
sufficiently strong, most individuals present in the population will
have a genotype with a small mutational distance from the wildtype.
The mutations that leave $d$ constant
($\boldsymbol{e}_{\xi}=\boldsymbol{e}_l-\boldsymbol{e}_k$), and those
that decrease $d$ ($\boldsymbol{e}_{\xi}=-\boldsymbol{e}_k$), which in
the case of a decreasing fitness function are neutral and
advantageous, respectively, occur with rates proportional to $x_k$,
and therefore are small for individuals with small mutational
distances $x=\sum_{k=1}^{3}x_{k}=d/N$, which form the main part of the
population, whereas the mutations that increase $d$
($\boldsymbol{e}_{\xi}=\boldsymbol{e}_k$) happen with rates
proportional to $1-x$, which are of order $1$ for small
$\boldsymbol{x}$. Therefore, it is reasonable to approximate the
mutation rates of the DNA model by unidirectional mutation rates,
which is the well known infinite sites limit (see \cite{Kim69} or
\cite{Ewe79}, Ch. 8).

In the case of unidirectional mutation, the mutational distance space
can be divided into three domains with respect to each sequence
$\boldsymbol{d}$, namely the ancestral cone, the offspring cone and
the sibling domain. Here, all sequences that can mutate to
$\boldsymbol{d}$ lie in the ancestral cone $AC(\boldsymbol{d})$, all
sequences that $\boldsymbol{d}$ can mutate to lie in the offspring
cone $OC(\boldsymbol{d})$, whereas the sequences that are not
connected to $\boldsymbol{d}$ via a mutational path form the sibling
domain $SD(\boldsymbol{d})$, which is the remainder of the mutational
distance space.

The eigenvalue equation of $\boldsymbol{H}$ for unidirectional
mutation is given by
\begin{equation}
\label{a1.1} 
\left(\overline{r}-\lambda_{\boldsymbol{d}}\right)\,p_{\boldsymbol{d}}
\;=\;
\sum_{k} u^{+k}_{\boldsymbol{d}-\boldsymbol{e}_k}\,
p_{\boldsymbol{d}-\boldsymbol{e}_k}^{}
\quad \mbox{with}\quad 
\lambda_{\boldsymbol{d}}\;=\;
r_{\boldsymbol{d}}-g(\boldsymbol{x}_{\boldsymbol{d}})
\;=\;r_{\boldsymbol{d}} -\sum_{k} u^{+k}_{\boldsymbol{d}}\,,
\end{equation}
where the diagonal entries are the eigenvalues
$\lambda_{\boldsymbol{d}}$, because $\boldsymbol{H}$ is a lower
triangular matrix. In this case, the situation is particularly simple,
and we can directly infer some properties of the population
distribution $\boldsymbol{p}$, which allow us to show that the maximum
principle holds in this case.

Suppose there is one $\widehat{\boldsymbol{d}}$ such that
$p_{\widehat{\boldsymbol{d}}}>0$, but
$p_{\widehat{\boldsymbol{d}}-\boldsymbol{e}_k}=0$ for
$k\in\{1,2,3\}$. This can only happen if
$\overline{r}=\lambda_{\widehat{\boldsymbol{d}}}$. An evaluation of
Eq.~(\ref{a1.1}) for $\widehat{\boldsymbol{d}}-\boldsymbol{e}_k$ and for
other sequences in the ancestral cone yields that no sequence in the
ancestral cone can contribute to the population, i.e.,
$p_{\boldsymbol{d}}=0$ for all $\boldsymbol{d}\in
AC(\widehat{\boldsymbol{d}})$.

If there was a sequence $\boldsymbol{d}^+$ in the offspring cone of
$\widehat{\boldsymbol{d}}$ with
$\lambda_{\boldsymbol{d}^+}>\lambda_{\widehat{\boldsymbol{d}}}$, we would
get $p_{\boldsymbol{d}^+}<0$, which is a contradiction to the
condition $p_i\ge 0$.  Thus,
$\lambda_{\boldsymbol{d}}\le\lambda_{\widehat{\boldsymbol{d}}}$ for all
$\boldsymbol{d}\in OC(\widehat{\boldsymbol{d}})$, which corresponds to the
maximum principle (\ref{5.1}) as $\overline{r}=\max_{\boldsymbol{d}\in
OC(\widehat{\boldsymbol{d}})}\left[r(\boldsymbol{x}_{\boldsymbol{d}})-
g(\boldsymbol{x}_{\boldsymbol{d}})\right]$.  If now
$\lambda_{\boldsymbol{d}^+}=\lambda_{\widehat{\boldsymbol{d}}}=\overline{r}$,
we get $p_{\boldsymbol{d}}=0$ for all $\boldsymbol{d}\in
AC(\boldsymbol{d}^+)$, including $\widehat{\boldsymbol{d}}$, so that in
this case the offspring cone $OC(\boldsymbol{d}^+)$ spans the
population rather than $OC(\widehat{\boldsymbol{d}})$.  Evaluating the
eigenvalue equation for the sequences in the offspring cone, we get
$p_{\boldsymbol{d}}>0$ for all $\boldsymbol{d}\in
OC(\widehat{\boldsymbol{d}})$ or $\boldsymbol{d}\in OC(\boldsymbol{d}^+)$,
respectively.

All sequences in the sibling domain $SD(\widehat{\boldsymbol{d}})$ descend
originally from the ancestors of $\widehat{\boldsymbol{d}}$ which are not
present in the population. Thus, their frequencies must vanish, unless
there is a sequence $\boldsymbol{d}^+\in SD(\widehat{\boldsymbol{d}})$
with $\lambda_{\boldsymbol{d}^+}=\lambda_{\widehat{\boldsymbol{d}}}$. In
this case, if $p_{\boldsymbol{d}^+}>0$, the sequences in both
offspring cones have non-vanishing frequency $p_{\boldsymbol{d}}>0$
for all $\boldsymbol{d}\in OC(\widehat{\boldsymbol{d}})\cup
OC(\boldsymbol{d}^+)$, the frequencies of all other sequences vanish.

As the mutation matrix is not irreducible for unidirectional mutation,
the population distribution in equilibrium is not unique, but depends
on the initial conditions. The equilibrium with the highest mean
fitness is always assumed if the wildtype initially occurs with
non-zero frequency, or if we consider the limit of small, but
non-vanishing back mutations.  In this case, we have
$\overline{r}=\lambda_{\rm
max}=\sup_{\boldsymbol{d}}\left[r(\boldsymbol{x_{\boldsymbol{d}}})-
g(\boldsymbol{x_{\boldsymbol{d}}})\right]$.

If, however, the wildtype is not present in the initial population
$\boldsymbol{p}(t=0)$ and the mutation rates for constant or
decreasing $d$ are exactly zero, we have to consider only the part of
the mutational distance space that is spanned by the offspring cones
$\bigcup_{\boldsymbol{d}_{\rm in}}OC(\boldsymbol{d}_{\rm in})$ of all
initially present sequences $\boldsymbol{d}_{\rm in}$. Now, the mean
fitness assumes the highest possible value in this subspace
$\overline{r}=\sup_{\cup \,OC(\boldsymbol{d}_{\rm
in})}\left[r(\boldsymbol{x}_{\boldsymbol{d}})-
g(\boldsymbol{x}_{\boldsymbol{d}})\right]$, which is assumed for at
least one sequence $\boldsymbol{d}^+$.

If this maximum is unique, the equilibrium population is given by the
right eigenvector corresponding to this eigenvalue
$\lambda_{\boldsymbol{d}^+}$, which has non-zero entries only for the
sequences in the offspring cone $OC(\boldsymbol{d}^+)$.  The left
eigenvector corresponding to $\lambda_{\boldsymbol{d}^+}$ has non-zero
entries only for the ancestors of $\boldsymbol{d}^+$, so that the only
$\boldsymbol{d}$ with $a_{\boldsymbol{d}}\neq 0$ is
$\boldsymbol{d}^+$, and hence $\boldsymbol{d}^+=\widehat{\boldsymbol{d}}$
is the only ancestor. This yields Eq.~(\ref{5.3}).

An interesting case arises, however, if the maximum is not unique, but
attained at two sequences $\widehat{\boldsymbol{d}}$ and
$\boldsymbol{d}^+$ in the subspace under consideration. In this
degenerate case, the ancestral distribution cannot be obtained as
easily as shown above, as the left and right eigenvectors have no
non-zero overlap and thus it is not possible to normalise
$\boldsymbol{z}$ such that $\sum_{i}a_{i}=1$.

We now have to distinguish between two cases. If the sequences lie in
parent-offspring relation, i.e., $\boldsymbol{d}^+\in
OC(\widehat{\boldsymbol{d}})$, $\widehat{\boldsymbol{d}}$ is the single
ancestor, whereas the population is formed by the offspring cone of
$\boldsymbol{d}^+$, i.e., $p_{\boldsymbol{d}}>0$ if and only if
$\boldsymbol{d}\in OC(\boldsymbol{d}^+)$. Note that Eq.~(\ref{5.3})
still holds, although in this special case the only ancestor
$\widehat{\boldsymbol{d}}$ has zero frequency in the population.

If, however, $\widehat{\boldsymbol{d}}$ and $\boldsymbol{d}^+$ lie in
sibling relation to each other, the population is formed by the
unification of their offspring cones $OC(\widehat{\boldsymbol{d}})\cup
OC(\boldsymbol{d}^+)$; and $\widehat{\boldsymbol{d}}$ and
$\boldsymbol{d}^+$ both may have non-vanishing ancestral frequency,
which are then determined by the initial conditions.

\subsection{Linear model}

We now consider the second case where the maximum principle applies
exactly. Here, both the fitness function and the mutation rates depend
linearly on functions $y_k$ of the genotype components $x_k$, and thus
can be written as
\begin{equation}
\label{5.2.1}
r(\boldsymbol{x})\;=\;
r_0-\sum_k \alpha_k\,y_k(x_k) \quad \mbox{and} \quad 
u^{\pm \xi}(\boldsymbol{x})\;=\;
u^{\pm \xi}_0+\sum_k \beta^{\pm \xi}_k\, y_k(x_k)\, ,
\end{equation}  
with parameters $\alpha_k$ and $\beta^{\pm \xi}_k$.  This type of
model has been used, e.g., in \cite{vHB86} and \cite{GH02} in a two-state
version.

In this case, the
mean fitness can be obtained by maximisation over the three components
of $\boldsymbol{x}$, and the maximum principle in the form of
Eqs.~(\ref{5.1}) and (\ref{5.3}) holds true for $y_k(x_k)=x_k$. This
can be shown by a direct calculation starting from Eq.~(\ref{4.9}),
which is carried out in Appendix~\ref{proof for the linear model}.

\subsection{Infinite sequence length}\label{Infinite sequence length}

\begin{figure}[tb]
  \begin{center}
  \includegraphics[width=\textwidth]{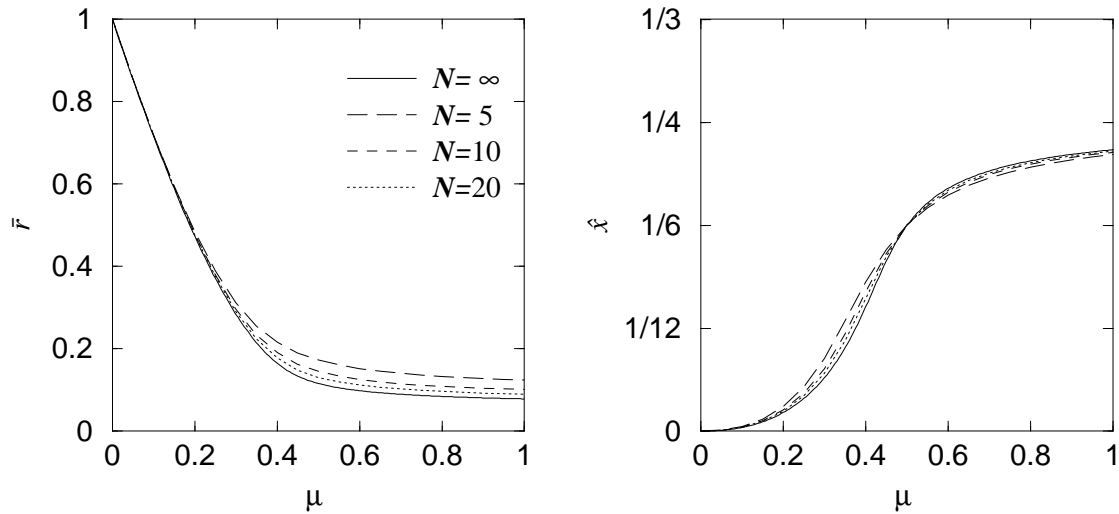}
  \caption{Population mean fitness $\overline{r}$ and ancestral mean 
mutational distance $\widehat{x}_k$ for varying mutation 
rate $\mu$ for a model with Jukes-Cantor mutation scheme and 
quadratic fitness $r=(1-\sum_{k=1}^3x_k)^2$. The curves for finite
sequence lengths $N=5,10,20$ are obtained by direct diagonalisation
of the time-evolution operator, the curve for $N=\infty$ is the result 
of the maximum principle.
}  
  \label{fig5}
  \end{center}
\end{figure}

\begin{figure}[tb]
  \begin{center}
  \includegraphics[width=\textwidth]{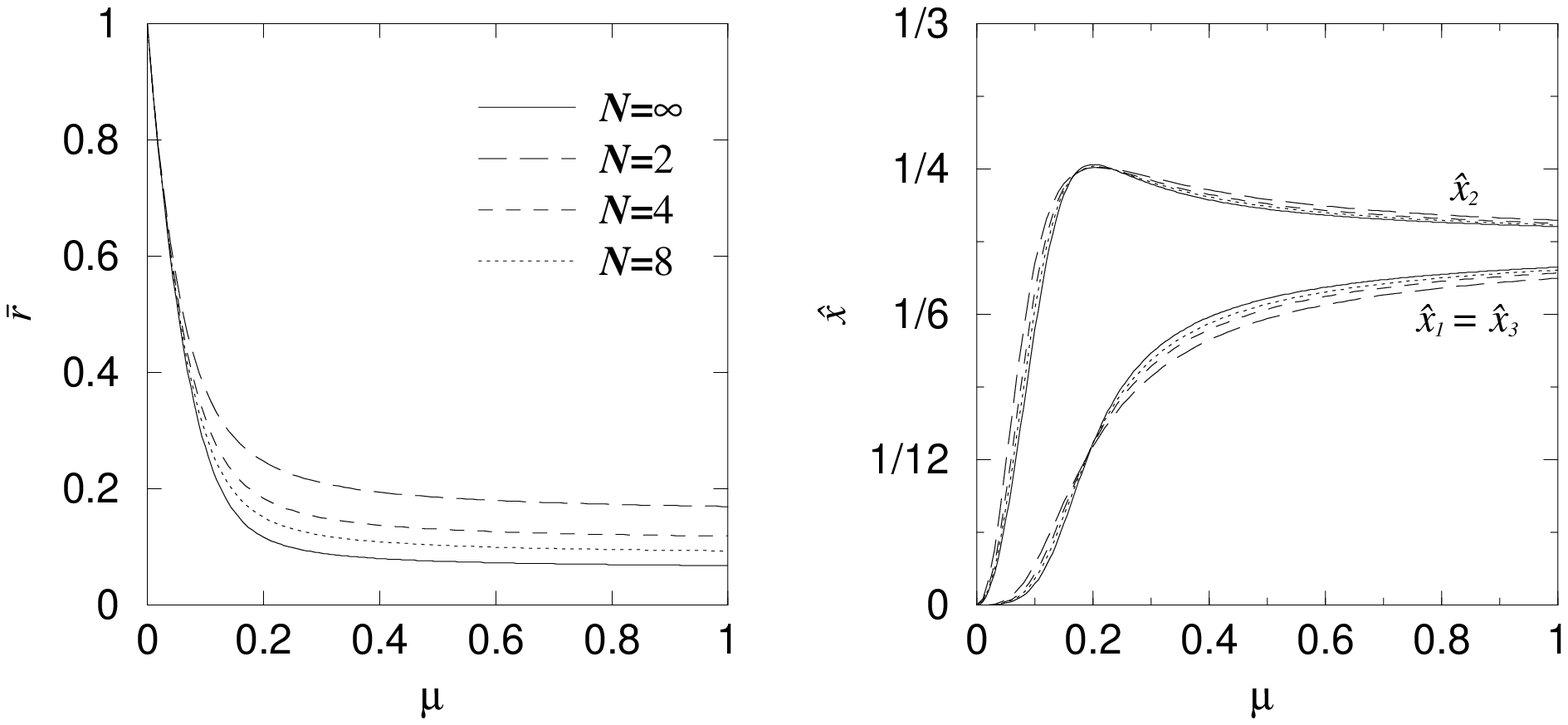}
  \caption{The quantities $\overline{r}$ and 
$\widehat{x}_k$ for varying mutation 
rate $\mu:=\mu_1=\mu_3$ for a model with K2P 
mutation scheme, with $\mu_2=10\mu$, and quadratic fitness 
$r=(1-\sum_{k=1}^3x_k)^2$. The curves for finite
sequence lengths $N=2,4,8$ are obtained by direct diagonalisation
of the time-evolution operator, the curve for $N=\infty$ is the result 
of the maximum principle.}  
  \label{fig6}
  \end{center}
\end{figure}

Finally, we consider the limit as the sequence length becomes
infinite.  In the limit as $N\rightarrow\infty$, we use
$\boldsymbol{x}_{\boldsymbol{d}}=\boldsymbol{d}/N$ to describe the
mutational distance to the wildtype. The quantities $x_{d_k}=d_k/N$
and $x_{\boldsymbol{d}}=\sum_{k=1}^3x_{d_k}$ fulfill the inequalities
$0 \le x_{d_k} \le 1$ and $0 \le x_{\boldsymbol{d}} \le 1$.  As long
as we operate with finite $N$, the $x_{d_k}$ take discrete rational
values only; in the limit $N\rightarrow\infty$, however, they become
dense, and it is thus reasonable to pass to a continuum formulation.
Consider the fitness function $r(\boldsymbol{x})$ and the mutation
rates $u^{\pm \xi}(\boldsymbol{x})$ as functions defined on the
mutational distance space. The mutation rates are positive, continuous
functions, obeying the boundary condition that they vanish for all
$\boldsymbol{x}$ at the boundary of the simplex in the mutational
distance space, where they correspond to mutations out of the set of
possible mutational distances. The fitness function has to be
piecewise continuous, i.e., it can have discontinuities only along
(finitely many) surfaces in the mutational distance space; at the
discontinuities, it must be either left or right continuous. This
allows for a number of biologically meaningful fitness functions,
like, for example, truncation selection.  For any finite $N$, the
fitness and mutation functions are sampled at all possible
$\boldsymbol{x}_{\boldsymbol{d}}$. The limit $N\rightarrow\infty$
is carried out such that the functions are kept constant, but the
sampling gets finer with increasing $N$.

In the limit as $N\rightarrow\infty$, Eqs.~(\ref{5.1}) and (\ref{5.3})
hold true. The proof is presented in Appendix~\ref{proof for infinite
sequence length}.  Even for rather short sequence lengths, the maximum
principle yields a reasonable approximation.

Figures \ref{fig5} and \ref{fig6} show numerical results obtained
using the maximum principle in comparison with those for finite
systems obtained by a direct diagonalisation of the time-evolution
operator.  For both figures, a quadratic symmetric fitness function
$r(\boldsymbol{x})=(1-\sum x_k)^2$ has been used, but they differ in
the mutation scheme. The mutation scheme used in Fig. \ref{fig5} is
the Jukes-Cantor mutation scheme, whereas in Fig. \ref{fig6}, we used
a Kimura 2 parameter (K2P) model with
$\mu_1=\mu_3:=\mu=\frac{1}{10}\mu_2$.  For the totally symmetric
Jukes-Cantor model, the ancestral means of the mutational distances
all coincide, whereas they differ in the K2P model, according to
$\widehat{x}_1=\widehat{x}_3<\widehat{x}_2$.

\section{Summary}

In this article, we investigated the mutation-selection balance in the
mutation-selection model introduced by \cite{HWB01}. There, a
deterministic approach to model the DNA evolution of asexual
populations was taken. We consider four-state sequences subject to the
forces of mutation and selection. For simplicity, selection is taken
to be permutation invariant, which leads to a three-dimensionally
structured mutational distance space, and the mutation model is a
single step model on this structure.

Using the concept of the ancestral distribution, as introduced by
\cite{HRWB02}, we derived a maximum principle for the population mean
fitness $\overline{r}$ and the ancestral mean genotype
$\widehat{\boldsymbol{x}}$ in equilibrium, which involves a maximisation
over the three dimensions of the mutational distance space.

This maximum principle gives the exact mean fitness in the three
limiting cases of (i) unidirectional mutation, (ii) linear fitness and
mutation functions, and (iii) the limit of infinite sequence
length. For finite sequence lengths, it is an approximation which we
expect to be correct up to terms of the order $1/N$. Numerically, we
found that already rather small sequence lengths were well reproduced.

The maximum principle generalises the results of \cite{HRWB02}, where
the case of a one-dimensional mutational distance space was treated,
which can be interpreted on the level of DNA sequences as a two-state
model, with states representing wildtype and mutant. In that case,
$\overline{r}$, $\widehat{x}$ and $\overline{x}=r^{-1}(\overline{r})$
could be obtained by a maximisation over one dimension.  In our model,
we have to maximise over the three dimensions of the mutational
distance space to obtain $\overline{r}$ and
$\widehat{\boldsymbol{x}}$, whereas the population mean genotype
$\overline{\boldsymbol{x}}$ cannot be derived as easily, because the
fitness function is not uniquely invertible in three dimensions.

Other quantities of interest are the corresponding variances. The
expressions for the variance of the fitness given by \cite{HRWB02} can
only be generalised to our model in the linear case, not for the case
of infinite sequence length. Neither can the variance of mutational
distance be obtained in a simply way, because this involves inversion
of the fitness function, which does not have a unique solution in more
than one dimension.

Although our setup is motivated by a model for DNA evolution, it is
valid for a system where the fitness depends on three arbitrary traits
$d_1$, $d_2$ and $d_3$ with a single-step mutation model, as long as
the mutation matrix describes a reversible process, i.e.,
$M_{ij}q_j=M_{ji}q_i$ with $\boldsymbol{q}$ being the equilibrium
distribution of the mutation process without selection.  In fact, the
maximum principle can be generalised to an $n$-state model, where we
have $n-1$ different traits determining the fitness. In this case, we
have to maximise over these $n-1$ quantities.

Similarly, the restriction to permutation-invariant fitness functions
could be dropped. In this case, however, we would have to maximise
over the $N$ sites of the sequence, so that it loses its use, which
lies primarily in the simplicity.

\begin{appendix}

\section{Proof of the maximum principle for the linear model} 
\label{proof for the linear model}

Starting with the eigenvalue equation of the symmetrised
time-evolution operator $\widetilde{\boldsymbol{H}}$ (\ref{4.9})
\begin{eqnarray}
\label{a2.1}
\overline{r} \sqrt{a_{\boldsymbol{d}}}&=&
\Big(r_{\boldsymbol{d}}
-\sum_{\xi}\big(u_{\boldsymbol{d}}^{+\xi}+
u_{\boldsymbol{d}}^{-\xi}\big)\Big)
\sqrt{a_{\boldsymbol{d}}} 
\;+\;
\sum_k\Big(
\sqrt{u^{+k}_{\boldsymbol{d}-\boldsymbol{e}_k}u^{-k}_{\boldsymbol{d}}}
\sqrt{a_{\boldsymbol{d}-\boldsymbol{e}_k}}
+\sqrt{u^{-k}_{\boldsymbol{d}+\boldsymbol{e}_k}u^{+k}_{\boldsymbol{d}}}
\sqrt{a_{\boldsymbol{d}+\boldsymbol{e}_k}}\Big) \nonumber\\
&&\;+\;\sum_{\substack{k,l \\ k\neq l}}
\sqrt{u^{+k-l}_{\boldsymbol{d}-\boldsymbol{e}_k+\boldsymbol{e}_l}
u^{-k+l}_{\boldsymbol{d}}}
\sqrt{a_{\boldsymbol{d}-\boldsymbol{e}_k+\boldsymbol{e}_l}} \,,
\end{eqnarray}
we make an ansatz for the $a_{\boldsymbol{d}}$ such that
\begin{equation}
\label{a2.2}
\frac{a_{\boldsymbol{d}-\boldsymbol{e}_k}}{a_{\boldsymbol{d}}}
\;=\;
C_k\,
\frac{u^{-k}_{\boldsymbol{d}}}{u^{+k}_{\boldsymbol{d}-\boldsymbol{e}_k}}\,.
\end{equation}
This is equivalent to
\begin{equation}
\label{a2.3}
\frac{a_{\boldsymbol{d}+\boldsymbol{e}_k}}{a_{\boldsymbol{d}}}
\;=\;
\frac{1}{C_k}\,
\frac{u^{+k}_{\boldsymbol{d}}}{u^{-k}_{\boldsymbol{d}+\boldsymbol{e}_k}}
\quad\mbox{and}\quad
\frac{a_{\boldsymbol{d}-\boldsymbol{e}_k+\boldsymbol{e}_l}}{a_{\boldsymbol{d}}}
\;=\;\frac{C_k}{C_l}\,
\frac{u^{-k+l}_{\boldsymbol{d}}}
{u^{+k-l}_{\boldsymbol{d}-\boldsymbol{e}_k+\boldsymbol{e}_l}} \,.
\end{equation}
The latter can be seen using the condition for reversibility
(\ref{4.6a}).  To determine the constants $C_k$, we multiply
Eq.~(\ref{a2.2}) by its denominators and sum over all
$\boldsymbol{d}$, which yields the ancestral means of the mutation
rates
\begin{equation}
\label{a2.4}
\sum_{\boldsymbol{d}} u^{+k}_{\boldsymbol{d}-\boldsymbol{e}_k}\,
a_{\boldsymbol{d}-\boldsymbol{e}_k}
\;=\; C_k \sum_{\boldsymbol{d}} u^{-k}_{\boldsymbol{d}}\, a_{\boldsymbol{d}}
\quad\Longleftrightarrow\quad 
C_k\; =\; \frac{\widehat{u}^{+k}}{\widehat{u}^{-k}} \,.
\end{equation}
Now, we divide Eq.~(\ref{a2.1}) by $\sqrt{a_{\boldsymbol{d}}}$, and insert
the ansatz (\ref{a2.2}) and (\ref{a2.3}),
\begin{equation}
\label{a2.5}
\overline{r}\; = \; 
r_{\boldsymbol{d}}
\;-\;
\sum_{\xi}\big(u^{+\xi}_{\boldsymbol{d}}+u^{-\xi}_{\boldsymbol{d}}\big)
\;+\;
\sum_k\Big(\sqrt{C_k}u^{-k}_{\boldsymbol{d}}+
\sqrt{\frac{1}{C_k}}u^{+k}_{\boldsymbol{d}}\Big)
\;+\;
\sum_{\substack{k,l \\ k\neq l}}
\sqrt{\frac{C_k}{C_l}}u^{-k+l}_{\boldsymbol{d}}\, .
\end{equation}
Multiplication by $a_{\boldsymbol{d}}$ and summation over all
$\boldsymbol{d}$ yields, using the explicit form (\ref{a2.4}) of the
$C_k$,
\begin{eqnarray}
\label{a2.6}
\overline{r} &=& \widehat{r}\;-\;
\sum_{\xi}\big(\widehat{u}^{+\xi}+\widehat{u}^{-\xi}\big)
\;+\;2\sum_k \sqrt{\widehat{u}^{+k}\widehat{u}^{-k}}
\;+\;\sum_{\substack{k,l \\ k\neq l}} 
\sqrt{\widehat{u}^{+k-l}\widehat{u}^{-k+l}} \nonumber \\
&=& \widehat{r}\;-\;\sum_{\xi} 
\Big(\widehat{u}^{+\xi}+\widehat{u}^{-\xi}-2\sqrt{\widehat{u}^{+\xi} 
\widehat{u}^{-\xi}}\,\Big)\,.
\end{eqnarray}
So far, we did not use linearity. If $r$ and $u$ depend linearly on
some functions $y_k(x_k)$, we have
$\widehat{r}=r(\widehat{\boldsymbol{y}})$ and $\widehat{u}^{\pm
\xi}=u^{\pm \xi}(\widehat{\boldsymbol{y}})$. With the definition of
the mutational loss function (\ref{5.2}) and in the case of
$y_k(x_k)=x_k$, this is Eq.~(\ref{5.3}).

In order to obtain the supremum condition (\ref{5.1}), we consider
Eq.~(\ref{a2.5}) for two different sequences $\boldsymbol{d}$ and
$\boldsymbol{d}'$ and take the difference, using the explicit
representation of fitness and mutation functions given in
Eq.~(\ref{5.2.1}),
\begin{eqnarray}
\label{a2.7}
0 &=& \sum_m \Bigg(
-\alpha_m\;-\;\sum_{\xi}\big(\beta^{+\xi}_m+\beta^{-\xi}_m\big)
\;+\;
\sum_k \Big(\, 
\sqrt{\frac{\widehat{u}^{+k}}{\widehat{u}^{-k}}}\, \beta^{-k}_m
+\sqrt{\frac{\widehat{u}^{-k}}{\widehat{u}^{+k}}}\, \beta^{+k}_m\, \Big)
\nonumber \\
&& \hphantom{\sum_m \Bigg(}
\;+\;
\sum_{\substack{k,l\\ k\neq l}} 
\sqrt{\frac{\widehat{u}^{+k-l}}{\widehat{u}^{-k+l}}}\, 
\beta^{-k+l}_m \Bigg)\,
\Big( y_m(x_m)-y_m(x_m') \Big) \,.
\end{eqnarray}
This is just the condition
\begin{equation}
\label{a2.8}
0\;=\;\sum_m\frac{\partial}{\partial y_m} 
\big[r(\boldsymbol{y})-g(\boldsymbol{y})\big]_{\boldsymbol{y}=
\widehat{\boldsymbol{y}}}\:
\big(y_m(x_m)-y_m(x_m')\big) \; ,
\end{equation}
which has to be fulfilled for arbitrary $\boldsymbol{x}$ and
$\boldsymbol{x}'$. Hence, we have
\begin{equation}
\label{a2.9}
0\; =\;\frac{\partial}{\partial y_m} 
\big[r(\boldsymbol{y})-g(\boldsymbol{y})\big]_{\boldsymbol{y}=
\widehat{\boldsymbol{y}}}
\qquad  \mbox{for $m\in\{1,2,3\}$.}
\end{equation}
This is a necessary condition for the existence of an extremum at
$\widehat{\boldsymbol{y}}$.  A sufficient condition for the existence
of a maximum of the function $r-g$ in $\widehat{\boldsymbol{y}}$ is
that the Hessian
\begin{equation}
\label{a2.9a}
\mathcal{H}_{mn}(\widehat{\boldsymbol{y}})\;:=\;\left[
\frac{\partial^2\big(r(\boldsymbol{y})-g(\boldsymbol{y})\big)}
{\partial y_m \partial y_n}
\right]_{\boldsymbol{y}=\widehat{\boldsymbol{y}}}
\end{equation}
of the second derivatives in the point $\widehat{\boldsymbol{y}}$ is a
negative definite matrix. We have
\begin{equation}
\label{a2.10}
\mathcal{H}(\boldsymbol{x})\;=\;-
\sum_z c_z^{}(\boldsymbol{x})\,
\boldsymbol{U}_z^{}(\boldsymbol{x})\,
\boldsymbol{U}_z^t(\boldsymbol{x})\, ,
\end{equation}
with $c_z(\boldsymbol{x})=\frac{1}{2}
\left(u^{+z}(\boldsymbol{x})u^{-z}(\boldsymbol{x})\right)^{-3/2}$,
$\boldsymbol{U}_z^t(\boldsymbol{x})=
\big(U_{z,1}(\boldsymbol{x}),U_{z,2}(\boldsymbol{x}),
U_{z,3}(\boldsymbol{x})\big)$, and $U_{z,m}(\boldsymbol{x})=\linebreak
\beta_m^{+z}u^{-z}(\boldsymbol{x})-\beta_m^{-z}u^{+z}(\boldsymbol{x})$.
To test the Hessian for negative definiteness, we evaluate the
quadratic form for an arbitrary vector $\boldsymbol{w}$. We have
\begin{equation}
\label{a2.11}
\boldsymbol{w}^t \mathcal{H} \boldsymbol{w}
\;=\;
-\sum_z c_z(\boldsymbol{x})\,
\big(w_1U_{z,1}(\boldsymbol{x})+w_2U_{z,2}(\boldsymbol{x})+
w_3U_{z,3}(\boldsymbol{x})\big)^2\,,
\end{equation}
which is $\le 0$ for all $\boldsymbol{w}$, and generically negative
unless all terms in the sum vanish.

Hence, there is a maximum at $\widehat{\boldsymbol{y}}$, and, together
with Eq.~(\ref{a2.6}), we have
\begin{equation}
\label{a2.12}
\overline{r}\;=\;\sup_{\boldsymbol{x}}
\big[r(\boldsymbol{y}(\boldsymbol{x}))-
g(\boldsymbol{y}(\boldsymbol{x}))\big]\;=\;
r(\widehat{\boldsymbol{y}})-g(\widehat{\boldsymbol{y}}) \,.
\end{equation}
In the case $y_k(x_k)=x_k$, this is the maximum principle as stated in
Eq.~(\ref{5.1}).

\section{Proof of the maximum principle for infinite sequence length}
\label{proof for infinite sequence length}

The proof of the maximum principle in the case $N\rightarrow\infty$
closely follows the corresponding proof in the two-state model,
compare \cite{HRWB02}.  The idea is to establish upper and lower
bounds for a system with finite $N$, which can be shown to converge
towards each other in the limit as $N\rightarrow\infty$.

In order to obtain a lower bound, we look at the system locally. To be
specific, we consider a volume $V_{s,\boldsymbol{d}_0}$ in the
mutational distance space around $\boldsymbol{d}_0$, containing the
sequences that can be reached from sequence $\boldsymbol{d}_0$ in at
most $s$ mutational steps. If this volume intersects a region where
$r$ has a jump, we take as $V_{s,\boldsymbol{d}_0}$ only that part
containing $\boldsymbol{d}_0$ where $r$ is continuous.  The number of
sequences contained in $V_{s,\boldsymbol{d}_0}$ is denoted by
$n(V_{s,\boldsymbol{d}_0})$.

The part of the time-evolution operator associated with this volume
consists of those matrix elements where both the row and the column
index correspond to sequences in $V_{s,\boldsymbol{d}_0}$. This
$n(V_{s,\boldsymbol{d}_0})\times
n(V_{s,\boldsymbol{d}_0})$-dimensional submatrix describes a system
with an effective mutational outflow, and hence the local growth rate
is lower than the global growth rate. Thus, the largest eigenvalue
$\overline{r}_{s,\boldsymbol{d}_0}$ of this submatrix yields a lower
bound for the largest eigenvalue $\overline{r}_N$ of the whole (but
still finite) system.

In order to obtain estimates for $\overline{r}_{s,\boldsymbol{d}_0}$,
we use the symmetrised system described by
$\widetilde{\boldsymbol{H}}$. This can be done because the eigenvalues
of the corresponding submatrices are the same.

We evaluate Rayleigh's principle for the quadratic form for the vector
$\boldsymbol{y}^t=(1,1,...,1)$ and get as lower bound for the largest
eigenvalue of the whole system
\begin{equation}
\label{a3.1}
\overline{r}_N
\;\ge\; \overline{r}_{s,\boldsymbol{d}_0}
\;=\;\sup_{\boldsymbol{y}}
\frac{\boldsymbol{y}^t\boldsymbol{Hy}}{\boldsymbol{y}^t\boldsymbol{y}}
\;\ge\;
\frac{\sum_{ij}(\widetilde{\boldsymbol{H}}_{s,\boldsymbol{d}_0})_{ij}}
{n(V_{s,\boldsymbol{d}_0})}\;.
\end{equation} 
To write this more explicitly, but in a compact way, we introduce the
function
\begin{equation}
\label{a3.2}
g_{N,\boldsymbol{d}}\;=\;
\sum_{\xi}\Big(u^{+\xi}_{\boldsymbol{d}}+u^{-\xi}_{\boldsymbol{d}}
-\sqrt{u^{+\xi}_{\boldsymbol{d}-\boldsymbol{e}_\xi}
u^{-\xi}_{\boldsymbol{d}}}
-\sqrt{u^{-\xi}_{\boldsymbol{d}+\boldsymbol{e}_{\xi}}
u^{+\xi}_{\boldsymbol{d}}}
\,\Big)\,,
\end{equation}
which sums over all mutational terms in the row labelled by
$\boldsymbol{d}$ of the full matrix $\widetilde{\boldsymbol{H}}$.
Using this, we get as a lower bound
\begin{equation}
\label{a3.3}
\overline{r}_N\ge \overline{r}_{s,\boldsymbol{d}_0}\;\ge\;
\frac{1}{n(V_{s,\boldsymbol{d}_0})} \left(\sum_{\boldsymbol{d}\in
V_{s,\boldsymbol{d}_0}}
\left(r_{\boldsymbol{d}}-g_{N,\boldsymbol{d}}\right)\;+\;
\sum\left(\mbox{boundary terms}\right)\right)\,.
\end{equation}
The boundary terms are the terms describing the mutational flow
through the surface $S_{s,\boldsymbol{d}_0}$ of the volume
$V_{s,\boldsymbol{d}_0}$ into $V_{s,\boldsymbol{d}_0}$ from the
outside, which are contained in the full $\widetilde{\boldsymbol{H}}$,
but not in $\widetilde{\boldsymbol{H}}_{s,\boldsymbol{d}_0}$. They are
of the form $\sqrt{u^{+\xi}_{\boldsymbol{d}-\boldsymbol{e}_{\xi}}
u^{-\xi}_{\boldsymbol{d}}}$ with $\boldsymbol{d}\in
V_{s,\boldsymbol{d}_0}$ and
$\boldsymbol{d}+\boldsymbol{e}_{\xi}\not\in V_{s,\boldsymbol{d}_0}$.

To perform the limit $N\rightarrow\infty$ as described in section
\ref{Infinite sequence length}, let
$r_{\boldsymbol{d}}=r(\boldsymbol{x}_{\boldsymbol{d}})$ and $u^{\pm
\xi}_{\boldsymbol{d}}=u^{\pm \xi}(\boldsymbol{x}_{\boldsymbol{d}})$ be
given as continuous functions, and analogously,
$g_{N,\boldsymbol{d}}=g_N(\boldsymbol{x}_{\boldsymbol{d}})$. The size
of $V_{s,\boldsymbol{d}_0}$ shall be scaled such that
$s_N\sim\sqrt{N}$.  With increasing $N$, the mutational distances
$\boldsymbol{x}$ of neighbouring sequences approach each other, and so
do the values of the functions $r$, $u^{\pm \xi}$ and $g_N$ as they
are continuous functions.  More precisely, the total Hamming distance
$\sum_{k=1}^3\left|d_k-d'_k\right|$ between any two sequences
$\boldsymbol{d},\boldsymbol{d}' \in V_{s,\boldsymbol{d}_0}$ is at most
$2s$. For the mutational distances we then have
$\boldsymbol{x}_{\boldsymbol{d}}-\boldsymbol{x}_{\boldsymbol{d}'}=
\frac{\boldsymbol{d}-\boldsymbol{d}'}{N}\rightarrow 0$ with increasing
$N$ because $\sum_k\frac{\left|d_k-d'_k\right|}{N}\le\frac{2s}{N}\sim
N^{-1/2}\rightarrow 0$.  For every $\boldsymbol{x}$ in the mutational
distance space, we can choose a suitable sequence
$(\boldsymbol{d}_N)=(\boldsymbol{d}_N(x))$ such that
$x_{\boldsymbol{d}_N}\rightarrow x$.  For any distance
$\boldsymbol{d}'$, such that $\boldsymbol{d}_N-\boldsymbol{d}'$ lies
in $V_{s_N,\boldsymbol{d}_N}$, we then have
$\lim_{N\rightarrow\infty}\left[
r(\boldsymbol{x}_{\boldsymbol{d}_N+\boldsymbol{d}'})-
g_N(\boldsymbol{x}_{\boldsymbol{d}_N+\boldsymbol{d}'})\right]=
r(\boldsymbol{x})-g(\boldsymbol{x})$.

On the other hand, the number of sequences in the volume
$V_{s_N,\boldsymbol{d}_0}$ increases with $N^{3/2}$, whereas the
number of sequences in the surface $S_{s_N,\boldsymbol{d}_0}$ of the
volume, and likewise the number of surface terms in Eq.~(\ref{a3.4}),
only increases with $N$. Therefore, we get
$\lim_{N\rightarrow\infty}\overline{r}_{s_N,\boldsymbol{d}_N}\ge
r(\boldsymbol{x})-g(\boldsymbol{x})$ for arbitrary $\boldsymbol{x}$.

For an upper bound, we consider a global maximum of the ancestral
distribution, i.e., a $\boldsymbol{d}^+$ such that
$a_{\boldsymbol{d}^+}\ge a_{\boldsymbol{d}}$ for all
$\boldsymbol{d}$. An evaluation of Eq.~(\ref{a2.1}) for
$\boldsymbol{d}^+$ and
$\sqrt{a_{\boldsymbol{d}'}}\le\sqrt{a_{\boldsymbol{d}^+}}$ yields
\begin{equation}
\label{a3.4}
\overline{r}_N\;\le\; r_{\boldsymbol{d}^+}-g_{N,\boldsymbol{d}^+}\;\le\; 
\sup_{\boldsymbol{d}}\big(r_{\boldsymbol{d}}-g_{N,\boldsymbol{d}}\big)\,.
\end{equation}
Performing the limit in the same way as above, we get
$r_{\boldsymbol{d}^+}-g_{N,\boldsymbol{d}^+}\rightarrow
r(\boldsymbol{x}^+)-g(\boldsymbol{x}^+)$, and combining this with the lower
bound, we have
\begin{equation}
\label{a3.5}
\sup_{\boldsymbol{x}}\big[r(\boldsymbol{x})-g(\boldsymbol{x})\big]
\;\le\; \overline{r}_{\infty}
\;\le\; r(\boldsymbol{x}^+)-g(\boldsymbol{x}^+)
\;\le\; \sup_{\boldsymbol{x}}
\big[r(\boldsymbol{x})-g(\boldsymbol{x})\big]\,,
\end{equation}
which proves Eq.~(\ref{5.1}).

Now, it remains to be shown that the ancestral distribution is peaked
around $\boldsymbol{x}^+$, and that indeed
$\boldsymbol{x}^+=\widehat{\boldsymbol{x}}$ as well as
$\widehat{r}=r(\widehat{\boldsymbol{x}})$. For this, we start again
with the eigenvalue equation in ancestral form (\ref{a2.1}), multiply
by $\sqrt{a_{\boldsymbol{d}}}$ and sum over the mutational distance
space
\begin{eqnarray}
\overline{r}_N &=& \sum_{\boldsymbol{d}}\Bigg[\Big(r_{\boldsymbol{d}}
-\sum_{\xi}\big(u^{+\xi}_{\boldsymbol{d}}+
u^{-\xi}_{\boldsymbol{d}}\big)\Big)\,
a_{\boldsymbol{d}}\nonumber\\
&&\hphantom{\sum_{\boldsymbol{d}}\Bigg[} \;+\;
\sum_{\xi}\Big(
\sqrt{u^{+\xi}_{\boldsymbol{d}-\boldsymbol{e}_{\xi}}
u^{-\xi}_{\boldsymbol{d}}}
\sqrt{a_{\boldsymbol{d}-\boldsymbol{e}_{\xi}}a_{\boldsymbol{d}}}
+\sqrt{u^{-\xi}_{\boldsymbol{d}+\boldsymbol{e}_{\xi}}
u^{+\xi}_{\boldsymbol{d}}}
\sqrt{a_{\boldsymbol{d}+\boldsymbol{e}_{\xi}}
a_{\boldsymbol{d}}}\Big)\Bigg] \,. 
\label{a3.6}
\end{eqnarray}
Using $\sqrt{\vphantom{h}a_{\boldsymbol{d}\pm\boldsymbol{e}_{\xi}}
a_{\boldsymbol{d}}}\le
\frac{1}{2}\left(a_{\boldsymbol{d}\pm\boldsymbol{e}_{\xi}}+
a_{\boldsymbol{d}}\right)$, we get
\begin{equation}
\label{a3.7}
\overline{r}_N\;\le\;
\sum_{\boldsymbol{d}}\big(r_{\boldsymbol{d}}-g_{N,\boldsymbol{d}}\big)\,
a_{\boldsymbol{d}}\;.
\end{equation}
As $\overline{r}_N\rightarrow\overline{r}_{\infty}$ and
$g_{N,\boldsymbol{d}}\rightarrow g(x_{\boldsymbol{d}})$ uniformly, for every
$\epsilon>0$ we can find an $N_{\epsilon}$ such that for every
$N>N_{\epsilon}$
\begin{equation}
\label{a3.8}
\overline{r}_{\infty}-\epsilon^2 \;<\; \sum_{\boldsymbol{d}}
\big(r(x_{\boldsymbol{d}})-
g(x_{\boldsymbol{d}})\big)\,a_{\boldsymbol{d}} \;.
\end{equation}
Due to Eq.~(\ref{a3.5}), we have
$r(x_{\boldsymbol{d}})-g(x_{\boldsymbol{d}})\le\overline{r}_{\infty}$.
Splitting the sum into two parts, $\sum_{\boldsymbol{d}_>}+
\sum_{\boldsymbol{d}_{\le}}$ with
$r(x_{\boldsymbol{d}_>})-g(x_{\boldsymbol{d}_>})>
\overline{r}_{\infty}-\epsilon$ and
$r(x_{\boldsymbol{d}_{\le}})-g(x_{\boldsymbol{d}_{\le}})\le
\overline{r}_{\infty}-\epsilon$, yields
\begin{equation}
\label{a3.9}
\overline{r}_{\infty}-\epsilon^2 \;<\;
\overline{r}_{\infty}\sum_{\boldsymbol{d}_>}a_{\boldsymbol{d}_>}
+(\overline{r}_{\infty}-\epsilon)
\sum_{\boldsymbol{d}_{\le}}a_{\boldsymbol{d}_\le}
\;=\;\overline{r}_\infty-\epsilon
\sum_{\boldsymbol{d}_{\le}}a_{\boldsymbol{d}_\le} \,.
\end{equation}
Thus, we have
$\sum_{\boldsymbol{d}_{\le}}a_{\boldsymbol{d}_\le}<\epsilon$, which
means that for $N$ sufficiently large, only sequences with
$r(\boldsymbol{x})-g(\boldsymbol{x})$ arbitrarily close to its maximum
$\boldsymbol{x}^+$ contribute to ancestral means.  Thus, in the
generic case that the maximum is unique, the ancestral distribution is
peaked around $\boldsymbol{x}^+=\widehat{\boldsymbol{x}}$, and thus
$\widehat{r}=r(\widehat{\boldsymbol{x}})$, which implies
Eq.~(\ref{5.3}).

\end{appendix}

\section*{Acknowledgements}
It is a pleasure to thank Ellen Baake, Michael Baake and Oliver Redner
for discussions and critical reading of the manuscript. We gratefully
acknowledge support by the British Council under the Academic Research
Collaboration (ARC) Programme, Project No 1213. We further wish to
express our gratitude to the Erwin Schr\"odiger International
Institute for Mathematical Physics in Vienna for support of several
extended stays in winter 2002/2003 in the programme ``Mathematical
Population Genetics and Statistical Physics''.  Finally, we thank an
anonymous referee for going to the time and effort of reading our
manuscript carefully and providing us with a list of detailed
suggestions for improvements.

\begin{small}

\end{small}


\begin{thebibliography}{36}
\expandafter\ifx\csname natexlab\endcsname\relax\def\natexlab#1{#1}\fi
\expandafter\ifx\csname url\endcsname\relax
  \def\url#1{{\tt #1}}\fi
\expandafter\ifx\csname urlprefix\endcsname\relax\def\urlprefix{URL }\fi

\bibitem[{Baake(1995)}]{Baa95}
Baake, E., 1995, Diploid models on sequence space, {\em Journal of Biological
  Systems\/} {\bf 3}, 343--349.

\bibitem[{Baake and Wagner(2001)}]{BW01}
Baake, E. and Wagner, H., 2001, Mutation-selection models solved exactly with
  methods of statistical mechanics, {\em Genetical Research\/} {\bf 78},
  93--117.

\bibitem[{Baake {\em et~al.\/}(1997)Baake, Baake, and Wagner}]{BBW97}
Baake, E., Baake, M., and Wagner, H., 1997, Ising quantum chain is equivalent
  to a model of biological evolution, {\em Physical Review Letters\/} {\bf 78},
  559--562, and Erratum, Physical Review Letters {\bf 79} (1997), 1782.

\bibitem[{Baake {\em et~al.\/}(2003)Baake, Baake, Bovier, and Klein}]{BBBK}
Baake, E., Baake, M., Bovier, A., and Klein, M., 2003, A simplified maximum
  principle for models of biological evolution (in preparation).

\bibitem[{Baxter(1982)}]{Bax82}
Baxter, R.~J., 1982, Exactly Solved Models in Statistical Mechanics, Academic
  Press, London.

\bibitem[{B\"urger(2000)}]{Bur00}
B\"urger, R., 2000, The Mathematical Theory of Selection, Recombination, and
  Mutation, Wiley, Chichester.

\bibitem[{Charlesworth(1990)}]{Cha90}
Charlesworth, B., 1990, Mutation-selection balance and the evolutionary
  advantage of sex and recombination, {\em Genetical Research Cambridge\/} {\bf
  55}, 199--221.

\bibitem[{Crow and Kimura(1970)}]{CK70}
Crow, J.~F. and Kimura, M., 1970, An Introduction to Population Genetics
  Theory, Harper \& Row, New York.

\bibitem[{Eigen(1971)}]{Eig71}
Eigen, M., 1971, Selforganization of matter and the evolution of biological
  macromolecules, {\em Naturwissenschaften\/} {\bf 58}, 465--523.

\bibitem[{Eigen {\em et~al.\/}(1989)Eigen, McCaskill, and Schuster}]{EMS89}
Eigen, M., McCaskill, J., and Schuster, P., 1989, The molecular quasi-species,
  {\em Advances in Chemical Physics\/} {\bf 75}, 149--263.

\bibitem[{Ewens(1979)}]{Ewe79}
Ewens, W.~J., 1979, Mathematical Population Genetics, Springer, Berlin.

\bibitem[{Ewens and Grant(2001)}]{EG01}
Ewens, W.~J. and Grant, G.~R., 2001, Statistical Methods in Bioinformatics,
  Springer, New York.

\bibitem[{Fisher(1922)}]{Fis22}
Fisher, R.~A., 1922, On the dominance ratio, {\em Proceedings of the Royal
  Society of Edinburgh\/} {\bf 42}, 321--341.

\bibitem[{Gerland and Hwa(2002)}]{GH02}
Gerland, U. and Hwa, T., 2002, On the selection and evolution of regulatory
  \mbox{DNA} motifs, {\em Journal of Molecular Evolution\/} {\bf 55}, 386--400.

\bibitem[{Haldane(1928)}]{Hal28}
Haldane, J. B.~S., 1928, A mathematical theory of natural and artificial
  selection. part v: Selection and mutation, {\em Proceedings of the Cambridge
  Philosophical Society\/} {\bf 23}, 838--844.

\bibitem[{Hamming(1950)}]{Ham50}
Hamming, R.~W., 1950, Error detecting and error correcting codes, {\em The Bell
  System Technical Journal\/} {\bf 26}, 147--160.

\bibitem[{Hartl and Clark(1997)}]{HC97}
Hartl, D.~L. and Clark, A.~G., 1997, Principles of Population Genetics,
  Sinauer, Sunderland, 3rd ed.

\bibitem[{Hermisson {\em et~al.\/}(2001)Hermisson, Wagner, and Baake}]{HWB01}
Hermisson, J., Wagner, H., and Baake, M., 2001, Four-state quantum chain as a
  model of sequence evolution, {\em Journal of Statistical Physics\/} {\bf
  102}, 315--343.

\bibitem[{Hermisson {\em et~al.\/}(2002)Hermisson, Redner, Wagner, and
  Baake}]{HRWB02}
Hermisson, J., Redner, O., Wagner, H., and Baake, E., 2002, Mutation selection
  balance: Ancestry, load, and maximum principle, {\em Theoretical Population
  Biology\/} {\bf 62}, 9--46.

\bibitem[{Jukes and Cantor(1969)}]{JC69}
Jukes, T.~H. and Cantor, C.~R., 1969, Evolution of protein molecules, in
  Mammalian Protein Metabolism (H.~N. Munro, ed.), Academic Press, New York,
  pp. 21--132.

\bibitem[{Kimura(1969)}]{Kim69}
Kimura, M., 1969, The number of heterozygous nucleotide sites maintained in a
  finite population due to steady flux of mutations, {\em Genetics\/} {\bf 61},
  893--903.

\bibitem[{Kimura(1980)}]{Kim80}
Kimura, M., 1980, A simple method for estimating evolutionary rate of base
  substitutions through comparative studies of nucleotide sequences, {\em
  Journal of Molecular Evolution\/} {\bf 16}, 111--120.

\bibitem[{Kimura(1981)}]{Kim81}
Kimura, M., 1981, Estimation of evolutionary distances between homologous
  nucleotide sequences, {\em Proceedings of the National Academy of Sciences
  USA\/} {\bf 78}, 454--458.

\bibitem[{Kohmoto {\em et~al.\/}(1981)Kohmoto, den Nijs, and Kadanoff}]{KNK81}
Kohmoto, M., den Nijs, M., and Kadanoff, L.~P., 1981, Hamiltonian studies of
  the $d=2$ \mbox{Ashkin--Teller} model, {\em Physical Review B\/} {\bf 24},
  5229--5241.

\bibitem[{Leuth\"ausser(1986)}]{Leu86}
Leuth\"ausser, I., 1986, An exact correspondence between \mbox{Eigen's}
  evolution model and a two-dimensional \mbox{Ising} system, {\em Journal of
  Chemical Physics\/} {\bf 84}, 1884--1885.

\bibitem[{Leuth\"ausser(1987)}]{Leu87}
Leuth\"ausser, I., 1987, Statistical mechanics of \mbox{Eigen's} evolution
  model, {\em Journal of Statistical Physics\/} {\bf 48}, 343--360.

\bibitem[{Li and Graur(1990)}]{LG90}
Li, W.~H. and Graur, D., 1990, Fundamentals of Molecular Evolution, Sinauer,
  Sunderland.

\bibitem[{O'Brien(1985)}]{OBr85}
O'Brien, P., 1985, A genetic model with mutation and selection, {\em
  Mathematical Biosciences\/} {\bf 73}, 239--251.

\bibitem[{Ohta and Kimura(1973)}]{OK73}
Ohta, T. and Kimura, M., 1973, A model of mutation appropriate to estimate the
  number of electrophoretically detectable alleles in a finite population, {\em
  Genetical Research\/} {\bf 22}, 201--204.

\bibitem[{Rumschitzky(1987)}]{Rum87}
Rumschitzky, D.~S., 1987, Spectral properties of \mbox{Eigen's} evolution
  matrices, {\em Journal of Mathematical Biology\/} {\bf 24}, 667--680.

\bibitem[{Swofford {\em et~al.\/}(1996)Swofford, Olsen, Waddell, and
  Hillis}]{SOWH96}
Swofford, D.~L., Olsen, G.~J., Waddell, P.~J., and Hillis, D.~M., 1996,
  Phylogenetic inference, in Molecular Systematics (D.~M. Hillis, C.~Moritz,
  and B.~K. Mable, eds.), Sinauer, Sunderland, pp. 407--514.

\bibitem[{Tarazona(1992)}]{Tar92}
Tarazona, P., 1992, Error thresholds for molecular quasispecies as phase
  transitions: From simple landscapes to spin-glass models, {\em Physical
  Review A\/} {\bf 45}, 6038--6050.

\bibitem[{van Lint(1982)}]{vLi82}
van Lint, J.~H., 1982, Introduction to Coding Theory, Springer, Berlin.

\bibitem[{von Hippel and Berg(1986)}]{vHB86}
von Hippel, P.~H. and Berg, O.~G., 1986, On the specificity of
  \mbox{DNA}--protein interactions, {\em Proceedings of the National Academy of
  Sciences USA\/} {\bf 83}, 1608--1612.

\bibitem[{Wiehe(1997)}]{Wie97}
Wiehe, T., 1997, Model dependency of error thresholds: The role of fitness
  functions and contrasts between finite and infinite sites models, {\em
  Genetical Research Cambridge\/} {\bf 69}, 127--136.

\bibitem[{Wright(1931)}]{Wri31}
Wright, S., 1931, Evolution in \mbox{Mendelian} populations, {\em Genetics\/}
  {\bf 16}, 97--159.

\end{thebibliography}
\end{document}